\documentclass[12pt]{article}

\usepackage{amsmath}
\usepackage{psfrag}
\usepackage[left=2cm,right=2cm,bottom=4cm, top=3cm]{geometry}

\usepackage{amsmath}
\usepackage{amsfonts}
\usepackage{mathtools}
\usepackage{comment}
\usepackage{color}
\usepackage[pdftex]{graphicx}
\usepackage{authblk}
\usepackage{bbold}
\usepackage{textcomp}
\usepackage{amssymb}
\usepackage{url}

\numberwithin{equation}{section}

\title{Optimal execution with nonlinear transient market impact}
\author[1]{Gianbiagio Curato}
\author[2]{Jim Gatheral}
\author[1,3,4]{Fabrizio Lillo}
\affil[1]{Scuola Normale Superiore, Piazza dei Cavalieri 7, 56126 Pisa, Italy}
\affil[2]{Baruch College, City University of New York, USA}
\affil[3]{Santa Fe Institute, 1399 Hyde Park Road, Santa Fe, NM 87501, USA}
\affil[4]{QUANTLab, Via Pietrasantina 123, 56122 Pisa, Italy}

\begin{document}

\maketitle

\begin{abstract}
We study the problem of the optimal execution of a large trade in the presence of nonlinear transient impact. We propose an approach based on homotopy analysis, whereby a well behaved initial strategy is continuously deformed to lower the expected execution cost. We find that the optimal solution is front loaded for concave impact and that its expected cost is significantly lower than that of conventional strategies. We then consider brute force numerical optimization of the cost functional; we find that the optimal solution for a buy program typically features a few short intense buying periods separated by long periods of weak selling.  Indeed, in some cases we find negative expected cost. We show that this undesirable characteristic of the nonlinear transient impact model may be mitigated either by introducing a bid-ask spread cost or by imposing convexity of the instantaneous market impact function for large trading rates.

\end{abstract}

\newpage

\tableofcontents

\newpage

\section{Introduction}

The optimization of trading strategies has long been an important goal for investors in financial markets. As demonstrated in the context of a linear equilibrium model by Kyle almost thirty years ago \cite{kyle}, the optimal strategy for an investor with insider information on the fundamental price of an asset is to trade incrementally through time. 
This strategy allows the trader to minimize costs whilst also minimizing the revelation of  information to the rest of the market. The precise way in which it is optimal to split the large order (herein called \textit{metaorder}) \cite{Zarinelli} depends on the objective function and on the market impact model, {\em i.e.} the change in price conditioned on signed trade size. 
In part due to the increasing tendency toward a full automation of exchanges and in part due to the discovery of new statistical regularities of the microstructure of financial markets, the problem of optimal execution is receiving growing attention from the academic and practitioner communities \cite{Abergel,Gatheral}.  

As pointed out in Gatheral et al. \cite{Gatheral2}, a first generation of market impact models \cite{Bertsimas,Almgren1,Almgren2,Almgren3} distinguishes between two impact components. The first component is temporary
and only affects the individual trade that has triggered it. The second component is permanent and affects all current and future trades equally. These models can be either in discrete or in continuous time and can assume either linear or nonlinear market impact for individual trades. The second generation of market impact models focusses on
the \textit{transient} nature of market impact \cite{Bouchaud,Obizhaeva,Bouchaud2,Gatheral}. In such models, market impact is typically assumed to factorize into two components:  instantaneous
market impact and a decay component. 
The instantaneous component models the reaction of price to traded volume. The decay component describes how the market price relaxes on average after the execution of an order.
In such models, each trade affects future price dynamics with an intensity that decays with  time.

The problem of optimal execution in the presence of transient impact has been considered in a series of recent studies. In the case of linear instantaneous market impact \cite{Gatheral2,Alfonsi2,Busseti}, the problem has been completely solved by showing that the cost minimization problem is equivalent to solving an integral equation. 
In particular Gatheral et al. \cite{Gatheral2} proved that optimal strategies can be characterized as measure-valued solutions of a Fredholm integral equation of the first kind. They show that optimal strategies always exist and are nonalternating between buy and sell trades when price impact decays as a convex function of time. 
This extends the result of Alfonsi et al. \cite{Alfonsi3} on the non existence of transaction triggered price manipulation, {\em i.e} strategies where the expected execution costs of a sell (buy) program can be decreased by intermediate buy (sell) trades.

However, a series of empirical studies \cite{Lillo,Bouchaud,Bacry} has clearly shown that the instantaneous market impact is a strongly concave function of the volume, well approximated by a power law function. The resulting optimal execution problem in the presence of nonlinear and transient impact is mathematically much more complicated than the linear case. In this paper we consider this optimization problem and propose several methods to find optimal solutions. 

Some important results in the nonlinear transient case were established by Gatheral \cite{Gatheral} who showed that under certain conditions,  the model admits price manipulation, {\em i.e.} the existence of round trip strategies with positive expected revenues. 
This money machine should of course be avoided in the modeling of market impact. In particular Gatheral set some necessary conditions for the absence of price manipulation (see below for details). 
A step toward the solution of the optimal execution problem under nonlinear transient impact has been made recently by Dang \cite{Dang}. In his paper, Dang suggests a way to convert the cost minimization problem into a nonlinear integral equation and proposes a numerical fixed point method on a discretization of the trading time interval to solve this equation. 
As we discuss in detail below, we find that Dang's fixed point method has convergence problems when the degree of nonlinearity of impact is significant and/or when the discretization grid is fine enough. 

In this paper we propose two methods to solve the optimal execution under nonlinear transient impact. The first is based on the Homotopy Analysis Method (HAM) \cite{Liao,Liao2,Liao3} applied to the discretized version of the integral equation proposed by Dang \cite{Dang}. The method starts from an initial guess and deforms it continuously in order to find better and better approximations of the solution of the integral equation. In doing this, we are implicitly restricting the space of solutions to continuous nonvanishing functions of the trading rate. We find that the optimal solution is a non time-symmetric U-shape; in the case of concave (convex) instantaneous impact, it is optimal to trade more at the beginning (end) of the metaorder. A comparative cost analysis shows that our solution outperforms conventional strategies. 

Once again, the HAM method explores only a restricted subspace of possible solutions. For this reason, in the second part of the paper we consider a fully numerical cost optimization method on a discrete grid. By using Sequential Quadratic Programming (SQP) we minimize directly the cost functional ({\em i.e.} we do not try to solve the integral equation). 
We find that the cost landscape is rugged, {\em i.e.} composed by a very large number of local minima separated by peaks. A significant number of these minima correspond to strategies with similar costs; for a buy program the corresponding strategy is an alternation of intense and short bursts of buying periods and long periods of weak selling. 
In other words, the model admits transaction triggered price manipulation. More important, when the nonlinearity is strong and/or the partition is fine, some such strategies have a negative expected cost, indicating that the model admits price manipulation. 
We then further extend our analysis by minimizing the cost functional with the extra constraint that all trades should have the same sign, so that for example selling is disallowed during execution of a buy metaorder. This case requires a derivative-free optimization method. By using a direct-search method, namely the Generating Set Search (GSS) method, we find positive expected
execution costs and \textit{sparse} optimal strategies, {\em i.e.} it is optimal to trade with a few bursts at a high trading rate interspersed with long periods of no-trading.    

In order to eliminate negative cost solutions, we propose two ways of regularizing the model, one based on the addition of a spread cost and one based on a modification of the instantaneous impact function. In the latter case the function becomes convex for sufficiently high trading rates. Both methods succeed in avoiding solution with negative costs and obviously reflect features of real markets.
  
The paper is organized as follows. In Section \ref{Gatheral's model and Fredholm integral equations}, we state the problem and explain why it is difficult to solve. We also briefly summarize our results on the convergence of the Dang fixed point method. In Section \ref{sec:Homotopy}, we present the HAM approach to the solution of the cost minimization problem and in Section \ref{sec:GeneralPiecewise} we present our results on the SQP and direct-search brute force minimization of the cost function. Section \ref{regularize} presents two proposed regularization methods.  In Section \ref{Conclusions}, we summarize and conclude.

\section{The optimal execution problem and its solution}
\label{Gatheral's model and Fredholm integral equations} 

The model of nonlinear transient market impact for the price $S\left(t\right)$ of an order execution starting at time $t=0$ when the price is $S(0)=S_0$ is 
\begin{equation}
 \label{eq:2.1}
S\left(t\right)=S_{0}+\int_{0}^{t} f\left(\dot x\left(s\right)\right)G\left(t-s\right)ds+\int_{0}^{t}\sigma\,dW(s),
\end{equation}
where $\dot x\left(s\right)$ is the rate of trading, {\em i.e.} number of shares per unit of time, at time $s<t$, $f\left(\dot x\left(s\right)\right)$ represents the impact of trading at time $s$, and $G\left(t-s\right)$
describes the impact decay. Finally, $\sigma$ is the volatility and $W(t)$ is a Wiener process. Thus $S\left(t\right)$ follows an arithmetic random walk with a drift that depends on the accumulated impacts of previous trades. We refer to $f\left(\cdot\right)$ as the \textit{instantaneous market impact function}
and to $G\left(\cdot\right)$ as the \textit{decay kernel}. In discrete time this is the propagator model originally developed by Bouchaud et al. \cite{Bouchaud,Bouchaud3}; the above continuous time formulation \eqref{eq:2.1} is due to Gatheral \cite{Gatheral}.  More recently, Bacry et al.\cite{Bacry} have shown how a market impact model of the form \eqref{eq:2.1} may be related to a more fundamental description of order flow using Hawkes processes.

The optimal execution problem consists in finding the trading strategy $\Pi=\lbrace {x}\left(t\right)\rbrace_{t \in [0,T]}$ that minimizes the execution cost for a given total amount $X$ of shares to be traded. The expected cost
$C\left[\Pi\right]$ associated with a given strategy $\Pi$ is given by
\begin{equation}
 \label{eq:2.2}
C\left[\Pi\right]=\mathbb{E}\left[\int_{0}^{T}\dot x\left(t\right)\left(S\left(t\right)-S_{0}\right)dt\right]=\int_{0}^{T}\dot x\left(t\right)\,\int_{0}^{t}f\left(\dot x\left(s\right)\right)G\left(t-s\right)ds\,dt,
\end{equation}
and the constraint that all shares should be traded is
\begin{equation}\label{constraint}
\int_0^T \dot x(t) ~ dt =X.
\end{equation}
Expression \eqref{eq:2.2} for the expected cost corresponds to expected implementation shortfall. We search for a statically optimal strategy. A statically optimal strategy is also dynamically
optimal when the cost depends on the stock price only trough the term $\int_{0}^{T}S\left(t\right)\dot x(t) ~ dt$, with $S\left(t\right)$ a martingale \cite{Predoiu}. Thus for the model of \eqref{eq:2.1} and the cost function described
by \eqref{eq:2.2}, a statically optimal strategy is also dynamically optimal.

The $\dot x\left(t\right)dt$ shares traded at time $t$ are traded at an expected price
\begin{equation}
 \label{eq:2.3}
S\left(t\right)=S_{0}+\int_{0}^{t}f\left(\dot x\left(s\right)\right)G\left(t-s\right)ds,
\end{equation}
which represents the cumulative impact of prior trading up to time $t$. 

The impact model of equation \eqref {eq:2.1} is fully specified by the form of the functions $f$ and $G$. A large body of empirical evidences points out two empirical facts on the form of these two functions. First, the instantaneous impact function $f(\cdot)$ is strongly concave. For example, based on a large sample of NYSE stocks, Lillo et al. \cite{Lillo} observed a concave function of the transaction volume. The concave function is well fitted by a power law with exponent 0.5 for small volumes and 0.2 for large volumes. Bouchaud et al. \cite{Bouchaud} analyzed stocks traded at the Paris Bourse and found that a logarithmic form gave the best fit to the data. In addition (in \cite{Bouchaud} for example),  the decay kernel $G(\cdot)$ is found to decay asymptotically as a power law function  
\begin{equation}
 \label{eq:2.5.1}
G(\tau) \sim \frac{1}{\tau ^\gamma}.
\end{equation}

The presence of these nonlinearities raises the question of the possible presence of price manipulation. There are different forms of manipulation. Following \cite{GatheralSchied}, we define a \textit{price manipulation} as a round trip trade  whose expected cost is negative. An impact model is free from price manipulation if, for any round trip trade, {\em i.e.} a strategy with $\int_{0}^{T}\dot x\left(t\right)dt=0$, the expected cost is non negative. According to Proposition 1 of \cite{GatheralSchiedSlynkoEcon}, the model of equation \eqref{eq:2.1} admits price manipulation in the nonlinear case unless the decay kernel $G\left(\tau\right)$ is singular for $\tau=0$.  For these reasons we will focus our analysis on decay kernels of the form $G\left(t-s\right)= \left(t-s\right)^{-\gamma}$. Moreover, as shown by Gatheral \cite{Gatheral}, the requirement of no price manipulation restricts the class of possible joint form for the instantaneous impact and decay kernel. Specifically, for power law impact function,  $f\left(\dot x\right)\propto sign\left(\dot x\right)|\dot x|^{\delta}$, and a power-law kernel, $G\left(t-s\right)= \left(t-s\right)^{-\gamma}$,  the following conditions  
\begin{eqnarray}
 \label{eq:2.4}
\ &&\gamma+\delta\geq1, ~~~~~~~~~~\gamma\geq\gamma^{*}=2-\frac{\log\,3}{\log\,2}\simeq0.415,
\end{eqnarray}
are {\it necessary} for the absence of price manipulations. In this paper we always consider parameters $\delta$ and $\gamma$ satisfying the above conditions. However, if these conditions are satisfied, there is no guarantee that the impact model does not admit price manipulation. Later in the paper we will show that it is indeed the case that the above conditions are not sufficient to preclude price manipulation.   A weaker form of price manipulation, relevant for our paper, is the {\em transaction-triggered} price manipulation defined in \cite{Alfonsi3}. This occurs when the expected revenues of a buy(sell) program may be increased by intermediate sell(buy) trades. 

\subsection{The case of linear market impact}

The optimization problem of minimizing the expected cost of equation \eqref{eq:2.2} under the constraint of equation \eqref{constraint} in the case of linear impact, $f(\dot x) \propto \dot x$, has been solved and widely studied \cite{Gatheral2}. In what follows, we use the symbol $v(t)$ to indicate the rate of trading $\dot x\left(t\right)$.  In particular, Proposition 22.9 of Ref. \cite{GatheralSchied}  states that if $G$ is positive definite, then $x(t)$ minimizes the expected cost if and only if there is a $\lambda$ such that $\forall t$, $x(t)$ solves
\begin{equation}\label{linearintegral}
\int_0^T G(|t-s|)dx(s)=\lambda
\end{equation}
As an important example, relevant for this paper, is the case $G\left(t-s\right)= \left(t-s\right)^{-\gamma}$ where the integral equation \eqref{linearintegral} becomes the Abel equation with solution
\begin{equation}\label{lineargss}
v\left( t \right)= \frac{c}{\left[t\,\left( T-t\right)\right]^{\frac{1-\gamma}{2}}},
\end{equation}
where $c$ is uniquely determined by the constraint equation \eqref{constraint} as 
\begin{equation}
 \label{eq:3.23.2}
c=X/\left(\sqrt{\pi}\left(\frac{T}{2}\right)^{\gamma}\frac{\Gamma\left(\left(1+\gamma\right)/2\right)}{\Gamma\left(1+\gamma/2\right)}\right),
\end{equation}
where $\Gamma\left(\cdot\right)$ is Euler's Gamma function. This solution is U-shaped and symmetric under time reversal, {\em i.e.} $ v(t) = v (T - t)$ , $t \in [0, T /2]$. In the following we will refer to this solution as the GSS solution. 

\subsection{The general case of nonlinear market impact}

In the general nonlinear case,  the problem is mathematically much more complicated. A first step in this direction has been presented very recently by Dang \cite{Dang} and consists in two contributions. 

The first one is the use of calculus of variations for integrals depending on convolution products \cite{Sanchez,Sanchez2,Sanchez3} to transform the cost minimization problem into an integral equation generalizing \eqref{linearintegral}. Specifically, given $f\in C^{1}\left(\mathbb{R}\right)$ and $G\in L^{1}\left[0,T\right]$, for the class of functions $x$ on $\left[0,T\right]$ satisfying
\begin{itemize}
 \item $x$ is absolutely continuous on $\left(0,T\right)$,
 \item $f\circ v\,\in\,L^{1}\left[0,T\right]$, 
\end{itemize}
the following necessary condition for the stationarity of the functional of equation \eqref{eq:2.2} holds:
\begin{equation}
 \label{eq:2.6}
\int_{0}^{t}f\left(v\left(s\right)\right)G\left(t-s\right)ds+f'\left(v\left(t\right)\right)\int_{t}^{T}v\left(s\right)G\left(s-t\right)ds=\lambda,
\end{equation}
where again $\lambda$ is a constant set by the constraint equation \eqref{constraint}.

In the case of a convex impact function, $f\left(v\right)\propto sign\left(v\right)|v|^{\delta}$ with $\delta>1$, equation \eqref{eq:2.6} holds $\forall v \in \mathbb{R}$.
In contrast, in the concave case, $\delta<1$, equation \eqref{eq:2.6} is not defined if the trading velocity $v$ vanishes at some time $t$, because the derivative of $f$ diverges at zero. This observation restricts the class of trajectories that can be considered. Moreover in the concave case there is no guarantee that the necessary condition \eqref{eq:2.6} is also sufficient, because the minimization cost problem could have a large number of extremal points

Equation \eqref{eq:2.6} is a weakly singular\footnote[1]{A kernel is called singular if it becomes infinite at one or more points in the range of integration, such as in the Abel's equation \cite{Wazwaz}.
A kernel is called weakly singular if its singularity is integrable, {\em i.e.} the integral of the function on a range that contains the singularity is finite. In our case the weakly singular kernel is given by $G\left(|t-s|\right)=|t-s|^{-\gamma}$.} Urysohn equation of the first kind \cite{Polyanin} taking the form 
\begin{equation}
 \label{eq:2.8}
\int_{0}^{T}G\left(|t-s|\right)\,F\left(v\left(s\right),t\right)\,ds= \lambda\,
\end{equation}
where
\begin{equation}
 \label{eq:2.9}
F\left(v\left(s\right),t\right)=\begin{cases} f\left(v\left(s\right)\right), & s\leq t \\ 
                                               v\left(s\right)f'\left(v\left(t\right)\right), & s>t.
          
         \end{cases}
\end{equation}

Note that there are two sources of nonlinearity in the integral equation \eqref{eq:2.8}: the nonlinear impact function $f\left(v\right)$, and the function $F$. In fact, the nonlinearity
depends also on the first derivative of the impact function $f'$, {\em i.e.} on the response of price to the traded volume per unit time. Moreover, the term involving $f'(v(t))$ entangles the response of price at time $t$ with the future trading rates, {\em i.e.} $v\left(s\right)$ for $s>t$, {\em i.e.} a coupling between present and future values of $v$. This means that equation \eqref{eq:2.8} cannot be classified as a weakly singular nonlinear Fredholm equation, because the function $F$ depends both on $t$ and on $s$.
In the linear impact case,  both nonlinearities disappear and one recovers a weakly singular linear Fredholm
integral equation of the first kind for the trading rate, where there is no such coupling between present and future times.

It is important to note that $F$ in equation \eqref{eq:2.8}  is not an invertible function of $v$, because it depends not only on $s$ but also on $t$. For this reason, the usual method for solving nonlinear integral equation by setting  $u(t)=F(v(t))$ and solving the linear integral equation for $u$ is not applicable here. In Section \ref{sec:Homotopy}, we use the Homotopy Analysis Method to solve the integral equation \eqref{eq:2.8}.


\subsubsection{Dang's fixed point algorithm}

The second contribution of Dang \cite{Dang} is a numerical scheme to solve the integral equation \eqref {eq:2.6}. This is a fixed point iteration scheme to find a numerical solution by quadrature methods which we review in Appendix \ref{sec:DangFixedPoint}.  Quadrature relies on a discretization of the time interval $[0,T]$ into $N$ subintervals, transforming the integral equation into a system of equations for a vector of dimension $N$. A detailed numerical analysis of the convergence properties of Dang's fixed point method as a function of the number of subintervals $N$ and the deviation of the impact from linearity, as measured by $\delta-1$ is presented in Appendix \ref{sec:DangFixedPoint}.   We now summarize the results of this analysis.

\begin{figure}[t]
	\centering
		\includegraphics[width=1.0\textwidth]{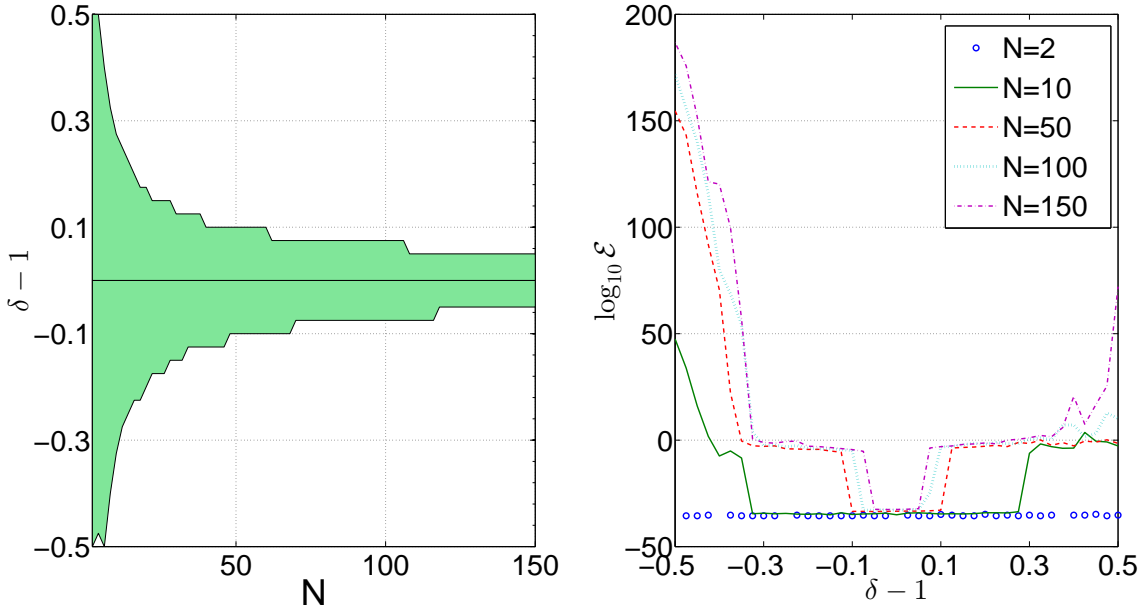} 
	\caption{Left panel: convergence region of the Dang's fixed point method on the parameter space $\left(N,\delta-1\right)$.
                 Right panel: squared residual error of solutions obtained as map's fixed points.}
	\label{fig:Conv_Region&Squared_Error}
\end{figure}

The green area described in the left panel of Figure \ref{fig:Conv_Region&Squared_Error} shows the region where our implementation of Dang's algorithm converges, while the right panel shows the squared residual error of solutions (defined in equation \eqref{eq:3.14} for a generic nonlinear operator) obtained as fixed points of the map. Both panels tell the same story. For small values of the number $N$ of subintervals, the method converges, even if the impact function is strongly nonlinear. As $N$ increases, {\em i.e} for finer partitions of the interval, we find that the method converges only for very moderately nonlinear impact functions. Dang's method obviously always converges when $\delta=1$, because in this case, the objective function has only one minimum. That Dang's method does not converge is no surprise; iteration schemes are useful for solving integral equations of the second kind, whilst equation \eqref{eq:2.8} is an integral equation of the first kind.

In summary, Dang's proposed solution technique in \cite{Dang} is problematic.  First, there is no guarantee that solving the integral equation \eqref{eq:2.6} leads to a strategy that minimizes the expected execution cost \eqref{eq:2.2} over the class of all admissible strategies. In particular , if the impact function is concave, as observed in reality, solutions are restricted to those for which the trading velocity $v(t)$ never vanishes, because $f'(v)$ diverges at $v=0$. Second, according to our numerical tests, Dang's fixed point algorithm seems to converge only for moderate nonlinearity and moderate discretizations of the $[0,T]$ interval.  Thus we need to investigate alternative approaches.

In the next Section we consider a perturbative approach to equation \eqref{eq:2.6} valid for weak nonlinearity, {\em i.e.} $\delta\simeq 1$.  This enables us to gain some understanding of the effect of nonlinearity on the optimal solution. In Section \ref{sec:Homotopy} we present a new approach to the solution of equation \eqref{eq:2.6} using the homotopy method. Finally, in Section \ref{sec:GeneralPiecewise},  we will present a brute-force numerical minimization of a discretized version of the cost function, where we will see explicitly that this cost function has many local minima. 

\subsection{A perturbative approach}
\label{Perturbative approach} 
In this section we present a simple perturbative method to investigate the solution of equation \eqref{eq:2.6} in the weakly nonlinear case. This will provide some intuition for the results we obtain later using the more powerful homotopy method. 
Our perturbative approach is based on two approximations. The first one regards the impact function, while the second one regards the trading rate, {\em i.e.} the unknown solution.  

Let us consider a buy program, $v(t)> 0$ and a slightly concave impact function, $f(v)=v^{1-\epsilon}$, with $0<\epsilon \ll 1$. We then make the approximations
\begin{equation}
 \label{eq:3.0.1}
f\left(v\right) = v-\epsilon \,v\,\log\left( v\right) + \mathcal{O}\left(\epsilon^2\right);\quad ~~~~f'\left( v\right) = 1-\epsilon -\epsilon\,\log\left( v\right) +\mathcal{O}\left( \epsilon^2 \right). 
\end{equation}
Substituting into equation \eqref{eq:2.6}, and keeping terms only to order $\epsilon$, we obtain
\begin{equation}
 \label{eq:3.0.2}
\begin{split}
& \int_{0}^{t}\,v\left( s\right)\,G\left( t-s\right)\,ds + \int_{t}^{T}\,v\left( s\right)\,G\left( s-t \right)\,ds \\
&-\epsilon\, \left\lbrace\int_{0}^{t}\,v\left( s\right)\log\left( v\left( s\right)\right)\,G\left( t-s\right)\,ds +\left[1+\log\left( v\left( t\right)\right)\right]\,\int_{t}^{T}\,v\left( s\right)\,G\left( s-t\right)\,ds\right\rbrace = \lambda.
\end{split}
\end{equation}
As mentioned earlier, the GSS solution \eqref{lineargss} solves the zeroth order case of  \eqref{eq:3.0.2}
\begin{equation}
 \label{eq:3.0.4}
\int_{0}^{t}\,v_0\left( s\right)\,G\left( t-s\right)\,ds + \int_{t}^{T}\,v_0\left( s\right)\,G\left( s-t\right)\,ds  = \lambda_0,
\end{equation}
in closed form.

\begin{figure}[t]
	\centering
		\includegraphics[width=0.8\textwidth]{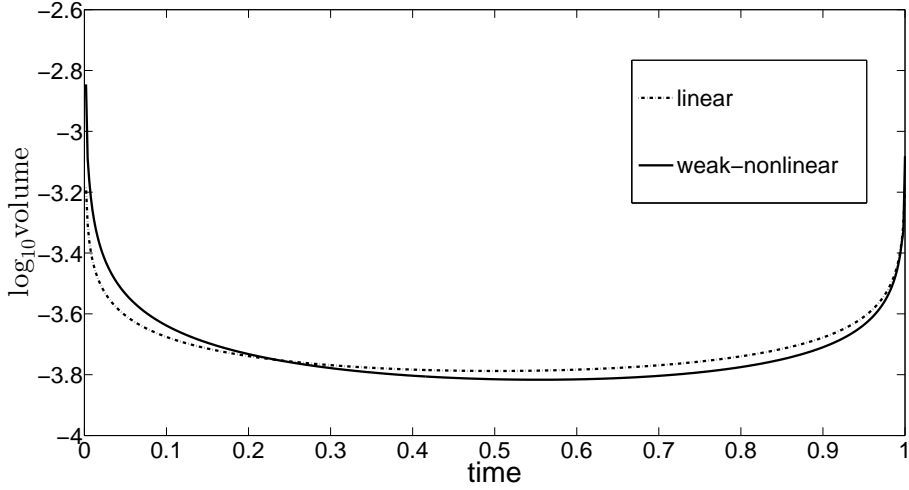} 
	\caption{Solution of the Urysohn equation in the weak nonlinear case for $\gamma=0.5,\epsilon=0.02$ and $X=0.1$. 
                 The full line represents the solution $v\left( s\right) = v_0\left( s\right) + \epsilon\,v_1\left( s\right)$.
                 We observe that this solution is not symmetric under time reversal.
                 The dotted line represents the GSS solution, {\em i.e.} the solution valid for the linear impact case.} 
	\label{fig:WeakCuratoGatheral}
\end{figure}

Now write $v(s) = v_0(s) +\epsilon \,v_1(s) + \mathcal{O}(\epsilon^2)$.  Then, matching terms of order $\epsilon$ gives the following linear Fredholm equation for $v_1\left(s\right)$:
\begin{eqnarray}
\label{eq:3.0.6}
&&\int_{0}^{t}\,v_1\left( s\right)\,G\left( t-s\right)\,ds + \int_{t}^{T}\,v_1\left( s\right)\,G\left( s-t\right)\,ds \nonumber\\
&&=\int_{0}^{t}\,v_0\left( s\right)\,\log\left( v_0\left( s\right)\right)\,G\left( t-s\right)\,ds +\left[1+\log\left( v_0\left( t \right)\right)\right]\,\int_{t}^{T}\,v_0\left( s\right)\,G\left( s-t\right)\,ds - \lambda'.\nonumber\\
\end{eqnarray}
We solve the above equation for a fixed value of $\lambda'$ using the constraint
\[
\int_{0}^{T}\,\left[v_0\left( s\right)+\epsilon \,v_1\left( s\right)\right]\,ds=X
\]
 to determine the correct value of $\lambda'$ for a fixed value of $\epsilon$.
We search for $\lambda'$ iteratively until equation \eqref{eq:3.0.6} is satisfied to within some given precision, in our case $1 \text{\textperthousand}$ of $X$. 


An example of this perturbative solution is shown in Figure \ref{fig:WeakCuratoGatheral}. 
We use the quadrature method described in Section \ref{sec:DiscreteHomotopy} to discretize the kernel $G$ and perform matrix inversion to solve a discretized version of equation \eqref{eq:3.0.6}. We report here the case $\gamma=0.5,\epsilon=0.02$ and $X=0.1$
 using a quadrature grid with $N=512$ points. 
Our first observation is that when market impact is concave, the optimal strategy is no longer symmetric under time reversal, in contrast to the linear case.  Indeed, it is better to trade faster in the first half of the trading period $T$ and more slowly in the second half. 
Repeating the computation in the convex market impact case, one obtains the opposite behavior: it is better to trade faster in the second half, as found also in \cite{Dang} with the fixed point method. In Section \ref{sec:Homotopy}, we will observe that the numerical solution of \eqref{eq:2.6} in the strongly nonlinear impact case using the homotopy method has the same properties.  It is worth emphasizing again that there is no reason to believe that a solution of the integral equation \eqref{eq:2.6} gives the optimal execution strategy, corresponding to the global minimum of the expected cost \eqref{eq:2.2}.


\section{The Homotopy Analysis Method}
\label{sec:Homotopy}
The concept of homotopy can be traced back to Henri Poincar\'{e}. In short, a homotopy describes a \textit{continuous} variation or deformation.
For example, a circle can be continuously deformed into a square or an ellipse, a coffee cup can be continuously deformed into a doughnut but not into a ball. More formally, let us consider the following
general nonlinear equation:
\begin{equation}
 \label{eq:3.1}
\mathcal{N}\left[v\left(t\right)\right]=0,
\end{equation}
where $\mathcal{N}$ is a nonlinear operator, $t$ denotes the independent variable, and $v\left(t\right)$ is the unknown function. Liao \cite{Liao} constructs the so-called zero-order deformation equation
\begin{equation}
 \label{eq:3.2}
\left(1-p\right) \mathcal{L}\left[\phi\left(t;p\right)-v^{0}\left(t\right)\right]=p\,\hslash\,H\left(t\right) \mathcal{N}\left[\phi\left(t;p\right)\right],
\end{equation}
where $p\in\left[0,1\right]$ is called the homotopy parameter, or embedding parameter and $\hslash$ is a non-zero auxiliary parameter which is called the convergence control parameter \cite{Abbasbandy2}. $H\left(t\right)\neq0$
is an auxiliary function, $\mathcal{L}$ is an auxiliary linear operator, $v^{0}\left(t\right)$ is an initial guess of $v\left(t\right)$, and $\phi\left(t;p\right)$ is an unknown function. 
There are no particular prescriptions for the choice of auxiliary functions or operators; often the choice depends on the problem  to be solved. When $p=0$ and $p=1$, we have respectively
\begin{equation}
 \label{eq:3.3}
\phi\left(t;0\right)=v^{0}\left(t\right),\,\,\,\,\, ~~~~~~~~~~\phi\left(t;1\right)=v\left(t\right).
\end{equation}
Thus, as $p$ increases from $0$ to $1$, the solution $\phi\left(t;p\right)$ varies \textit{continuously} from the initial guess $v^{0}\left(t\right)$ to the sought solution $v\left(t\right)$. Expanding $\phi\left(t;p\right)$
in a Maclaurin series with respect to $p$, we have
\begin{equation}
 \label{eq:3.4}
\phi\left(t;p\right)=v^{0}\left(t\right)+\sum_{m=1}^{\infty}v^{m}\left(t\right)p^{m},
\end{equation}
where
\begin{equation}
 \label{eq:3.5}
v^{m}\left(t\right)=\frac{1}{m!}\frac{\partial^{m} \phi\left(t;p\right)}{\partial p^{m}} \Big|_{p=0}.
\end{equation}
The series representation of $\phi$ in equation \eqref{eq:3.4} is called the homotopy series and $v^m(t)$ in \eqref{eq:3.5} is called the $m$th-order homotopy derivative of $\phi$ \cite{Liao3}. If the auxiliary linear operator, the initial guess, the convergence control
parameter $\hslash$, and the auxiliary function are properly chosen, the homotopy series converges at $p=1$, giving the sought solution of equation \eqref{eq:3.1}. Then by using the relationship $\phi\left(t;1\right)=v\left(t\right)$, one has the so-called homotopy series
solution
\begin{equation}
 \label{eq:3.6}
v\left(t\right)=v^{0}\left(t\right)+\sum_{m=1}^{\infty}v^{m}\left(t\right),
\end{equation}
which is one of the solutions of the original nonlinear equation, as proved by Liao \cite{Liao4}. Defining the vector 
\begin{equation}
 \label{eq:3.7}
\textbf{v}^{m}=\left\lbrace v^{0}\left(t\right),v^{1}\left(t\right),\cdots,v^{m}\left(t\right) \right\rbrace,
\end{equation}
and differentiating the zero-order deformation equation \eqref{eq:3.2} $m$ times with respect to the homotopy parameter $p$ and then setting $p=0$ and finally dividing them by $m!$, we have the so-called $m$th-order
deformation equation
\begin{equation}
 \label{eq:3.8}
\mathcal{L}\left[v^{m}\left(t\right)-\chi^{m}v^{m-1}\left(t\right)\right]=\hslash\, H\left(t\right) R^{m}\left(\textbf{v}^{m-1}\right),
\end{equation}
where
\begin{equation}
 \label{eq:3.9}
R^{m}\left(\textbf{v}^{m-1}\right)=\frac{1}{\left(m-1\right)!}\frac{\partial^{m-1} \mathcal{N}[\phi\left(t;p\right)]}{\partial p^{m-1}}\Big|_{p=0}
\end{equation}
and
\begin{equation}
 \label{eq:3.10}
\chi^{m}=\begin{cases} 0, & m\leq1\,, \\ 
                       1, & m>1.
          
         \end{cases}
\end{equation}
The $n$th-order approximate solution is then given by
\begin{equation}
 \label{eq:3.11}
v^{(n)}\left(t\right)=v^{0}\left(t\right)+\sum_{m=1}^{n} v^{m}\left(t\right),
\end{equation}
and the exact solution by $v\left(t\right)=\lim_{n\rightarrow \infty} v^{(n)}\left(t\right)$. The two main difficulties of this approach are to compute the derivatives of equation \eqref{eq:3.9} and how to
choose an appropriate value of $\hslash$ in order to guarantee the convergence of the series solution of equation \eqref{eq:3.6}. 

The first problem requires us to compute the homotopy derivative of a given nonlinear smooth function $f$ of the homotopy Maclaurin series of equation \eqref{eq:3.4}, {\em i.e.} to compute the derivative of $f\left(\phi\right)$. We refer the reader to Appendix \ref{appHomo} for details.  The second problem is how to choose an appropriate value of the convergence control parameter $\hslash$ in order to guarantee the convergence of the series of equation \eqref{eq:3.6} \cite{Liu}. 
To do this, we will adopt in the following, the so-called \textit{optimization} method \cite{Liao,Akyildiz,Abbasbandy5}, according to which, we define the squared residual of the governing equation \eqref{eq:3.1} as
\begin{equation}
 \label{eq:3.14}
\mathcal{E}^{n}\left(\hslash\right)=\int_{0}^{T}\left(\mathcal{N}\left[v^{(n)}\left(t\right)\right]\right)^2dt.
\end{equation}
The optimal value of the convergence control parameter is then obtained by finding the minimum of this squared residual. In fact, if $v^{(n)}(t)$ is the solution of the original problem of equation \eqref{eq:3.1}, the residual $\mathcal{E}^{n}\left(\hslash\right)$ vanishes.

\subsection{Homotopy for nonlinear transient market impact}
\label{HAM for a nonlinear market impact}    

Now we apply the Homotopy Analysis Method (HAM) to the solution of the nonlinear integral equation \eqref{eq:2.8}. Thus 
\begin{equation}
 \label{eq:3.16}
\mathcal{N}\left[v(t)\right]=-\lambda+\int_{0}^{T}G\left(|t-s|\right)\,F\left(v(s),t\right)\,ds.
\end{equation}

\noindent As suggested in \cite{Hetmaniok}, we choose the linear operator $ \mathcal{L}$ to be the identity, that is
\begin{equation}
 \label{eq:3.15}
 \mathcal{L}\left[\phi\left(t;p\right)\right]=\phi\left(t;p\right),
\end{equation}
and the auxiliary function to be $H\left(t\right)=1$. The zero-order deformation equation is then
\begin{equation}
 \label{eq:3.17}
\left(1-p\right)\left[\phi\left(t;p\right)-v^{0}\left(t\right)\right]=\hslash\,p\,\mathcal{N}\left[\phi\left(t;p\right)\right].
\end{equation}
Differentiating this zero-order deformation equation $m$ times with respect to $p$,
and finally dividing by $m!$, we obtain the $m$th-order deformation equation
\begin{equation}
 \label{eq:3.18}
v^{m}\left(t\right)=\chi^{m} v^{m-1}\left(t\right)+\hslash \,R^{m}\left(\textbf{v}^{m-1}\right),
\end{equation}
where for $m>1$
\begin{eqnarray}
\label{eq:3.19}
\ R^{m}\left(\textbf{v}^{m-1}\right)&=&\frac{1}{\left(m-1\right)!}\frac{\partial^{m-1} \mathcal{N}\left[\phi\left(t;p\right)\right]}{\partial p^{m-1}} |_{p=0} \nonumber\\
\                                &=&\int_{0}^{T}G\left(|t-s|\right)\left\lbrace\frac{1}{\left(m-1\right)!}\left.\frac{\partial^{m-1}F\left(\phi\left(s;p\right),t\right)}{\partial p^{m-1}}\right|_{p=0}\right\rbrace\,ds\,. \nonumber\\
\end{eqnarray}
For example, the first-order deformation equation is 
\begin{equation}
 \label{eq:3.20}
v^{1}\left(t\right)=\hslash\,\left(-\lambda+\int_{0}^{T}G\left(|t-s|\right)F\left(v^{0}\left(s\right),t\right)\,ds\right).
\end{equation}
To compute the higher-order deformation equations we need to write $F$ as a function of two homotopy-series, defined respectively at the time points $s$ and $t$
\begin{equation}
 \label{eq:3.21}
F\left(\sum_{i=0}^{\infty}v^{i}\left(s\right)p^{i},\sum_{j=0}^{\infty}v^{j}\left(t\right)p^{j}\right)=\begin{cases} f\left(\sum_{i=0}^{\infty}v^{i}\left(s\right)p^{i}\right), & s\leq t \\ 
                                                                                                                     \left(\sum_{i=0}^{\infty}v^{i}\left(s\right)p^{i}\right)\,f'\left(\sum_{j=0}^{\infty}v^{j}\left(t\right)p^{j}\right), & s>t.
          
         \end{cases}
\end{equation}  
and then apply the homotopy-derivative of equation \eqref{eq:3.12}, relative to a nonlinear one-dimensional system, for $s\leq t$, and the homotopy-derivative of equation \eqref{eq:3.13}, relative to a nonlinear two-dimensional system, for $s>t$. 
The expression of the first homotopy-derivative, for $v^{0}\left(t\right)>0$, is given by
\begin{equation}
 \label{eq:3.22}
\left.\frac{\partial}{\partial p} F\left(\sum_{i=0}^{\infty}v^{i}\left(s\right)p^{i},\sum_{j=0}^{\infty}v^{j}\left(t\right)p^{j}\right)\right|_{p=0}=\begin{cases}  \delta\left(v^{0}\left(s\right)\right)^{\delta-1}v^{1}\left(s\right), & s\leq t \\ 
                                                                                                                                                        \delta\, v^{1}\left(s\right)\left(v^{0}\left(t\right)\right)^{\delta-1}+\delta\left(\delta-1\right)v^{0}\left(s\right)\left(v^{0}\left(t\right)\right)^{\delta-2}v^{1}\left(t\right), & s>t.
          
         \end{cases}
\end{equation} 
Higher orders may be computed using symbolic computation software such as Mathematica or Maple. 

Obviously, for this algorithm to work, the initial guess needs to satisfy the condition $f'(v^0(t)) < \infty$ for all $t \in [0,T]$.  Thus, $v^{0}$ must have the same sign on the whole interval $[0,T]$. From now on, we consider a buy program.  In order to test for possible dependencies on the initial guess, we choose both the VWAP strategy, $v^{0}_{VWAP}\left(t\right)=X/T$, and the GSS solution for the linear case, {\em i.e.} $v^{0}_{GSS}$ given by equation \eqref{lineargss} as initial guesses. 

\subsection{A Discrete Homotopy Analysis Method}
\label{sec:DiscreteHomotopy}

To apply HAM to our problem, we need to compute the definite integrals \eqref{eq:3.19}, which seem to be analytically intractable.  We therefore propose a way to approximate these integrals, and
refer to this discretized version of HAM as the DHAM approach.

We first discretize equation \eqref{eq:2.8} by splitting the time interval $\left[0,T\right]$ into $N$ subintervals at the times $t_{i}=i\,T/N$ where $i\in \left\lbrace 0,1,\cdots,N \right\rbrace$. This gives the following nonlinear system of
$N$ equations in the variables $v_{i}=v\left(t_{i}\right)$ where $i\in \left\lbrace 1,\cdots,N \right\rbrace$
\begin{equation}
 \label{eq:3.24}
\sum_{j=1}^{N}\,G_{ij}F_{ij}\left(v\right)=\lambda
\end{equation}
where $i$ indicates the time point $t_{i}$.  The nonlinear function $F\left(.\right)$ of equation \eqref{eq:2.9} becomes a real $N\times N$ matrix
\begin{equation}
\label{eq:3.25}
 F_{ij}\left(v\right)=\begin{cases} f\left(v_{j}\right), & j\leq i \\ v_{j}f'\left(v_{i}\right), & j>i.
\end{cases}
\end{equation}
The decay kernel $G\left(|t-s|\right)$ becomes a Toeplitz real symmetric $N \times N$ matrix given by
\begin{equation}
 \label{eq:3.26}
G_{ij}=\int_{t_{i-1}}^{t_{i}}\,\int_{t_{j-1}}^{t_{j}}\,G\left(|t-s|\right)\,ds~dt.
\end{equation}
If $G\left(\tau \right)=\tau^{-\gamma}$, for $i>j$ we have
\begin{equation}
 \label{eq:3.27}
G_{ij}=\frac{1}{\left(1-\gamma\right)\left(2-\gamma\right)}\left(\frac{T}{N}\right)^{2-\gamma}\left\lbrace \left(i-j+1\right)^{2-\gamma}
-2\left(i-j\right)^{2-\gamma}+\left(i-j-1\right)^{2-\gamma} \right\rbrace
\end{equation}
and the diagonal terms are given by
\begin{equation}
 \label{eq:3.28}
G_{ii}=\frac{2}{\left(1-\gamma\right)\left(2-\gamma\right)}\left(\frac{T}{N}\right)^{2-\gamma}.
\end{equation}
In this scheme the constraint on the traded volume of equation \eqref{constraint} is given by
\begin{equation}
 \label{eq:3.29}
\sum_{i=1}^{N} v_{i}=\frac{N X}{T}.
\end{equation}
We are not able to solve the nonlinear system of equation \eqref{eq:3.24} directly.  Rather, we search an approximate homotopy series solution
\begin{equation}
 \label{eq:3.30}
 v_{i}=v^{0}_{i}+\sum_{m=1}^{\infty} v_{i}^{m},\,\,\,i\in\left\lbrace 1,\cdots,N \right\rbrace,
\end{equation}
where the deformation equations \eqref{eq:3.18} for $m>1$ are approximated by
\begin{equation}
 \label{eq:3.31}
v_{i}^{m}=v_{i}^{m-1}+\hslash \sum_{j=1}^{N} G_{ij}\,F^{m-1}_{ij},
\end{equation}
and where the $\left(m-1\right)$-th homotopy derivative is evaluated on the grid points $t_{i},\,s_{j}$
\begin{equation}
 \label{eq:3.32}
F^{m-1}_{ij}=\frac{\partial^{m-1}}{\partial p^{m-1}} F\left(\sum_{k=0}^{\infty}v_{i}^{k}\,p^{k},\sum_{l=0}^{\infty}v_{j}^{l}\,p^{l}\right)\Big|_{p=0}.
\end{equation}
The approximate solution of order $n$ is given by
\begin{equation}
 \label{eq:3.33}
v_{i}^{(n)}=v^{0}_{i}+\sum_{m=1}^{n} v_{i}^{m},
\end{equation}
and we compute the squared residual error of equation \eqref{eq:3.24} as
\begin{equation}
 \label{eq:3.34}
\mathcal{E}^{n}\left(\hslash\right)=\sum_{i=1}^{N}\left[-\lambda+\sum_{j=1}^{N}G_{ij}\,F_{ij}\left(v^{(n)}\right)\right]^2.
\end{equation}
The value of $\hslash_{min}$ that minimizes this error gives the DHAM solution $v^{(n)}_{i}\left(\hslash_{min}\right)$ of equation \eqref{eq:2.8}.
This solution can be considered as a piecewise constant approximation of the exact solution corresponding to a sequence
of VWAP executions with trading rates $v^{(n)}_{i}$. 

Finally, the expected liquidation cost  \eqref{eq:2.2} is approximated by
\begin{equation}
\label{eq:3.35}
C\left[v^{(n)}\right]= \sum_{i=1}^{N}\,\sum_{j=1}^{N} v_{i}^{(n)}\,f\left(v_{j}^{(n)}\,\right) A_{ij},
\end{equation}
where the $A_{ij}$ are elements of a Toeplitz matrix that describes the decay kernel $G\left(t-s\right)$
\begin{eqnarray}
\label{eq:3.36}
\  A_{ij}&=&0;\,\,j>i ,    \nonumber \\
\  A_{ii}&=&G_{ii}/2; \nonumber \\
\  A_{ij}&=&G_{ij};\,\,j\leq i. 
\end{eqnarray} 

\subsection{DHAM results}
\label{Results} 

We present here the optimal strategies obtained with DHAM in the no-dynamic-arbitrage region given by equation \eqref{eq:2.4}. 
We analyze in detail the values $\gamma=0.45$ and $\gamma=0.5$ in a strong nonlinear regime, i.e. $\delta=0.5$.
We consider the case where the volume to be purchased is $X=0.1$, which can be interpreted as a metaorder execution where one buys $10\%$ of the available unitary market volume. We discretize using a grid of $N=100$ subintervals and compute the solution up to the  $7$-th order with a tolerance on the constraint on the total quantity executed  of $1$\text{\textperthousand} of $X$. As mentioned earlier, we consider two initial guesses, namely one corresponding to a VWAP profile and the other to a GSS solution \eqref{lineargss}. 


\begin{figure}[th]
\includegraphics[width=1.0\textwidth]{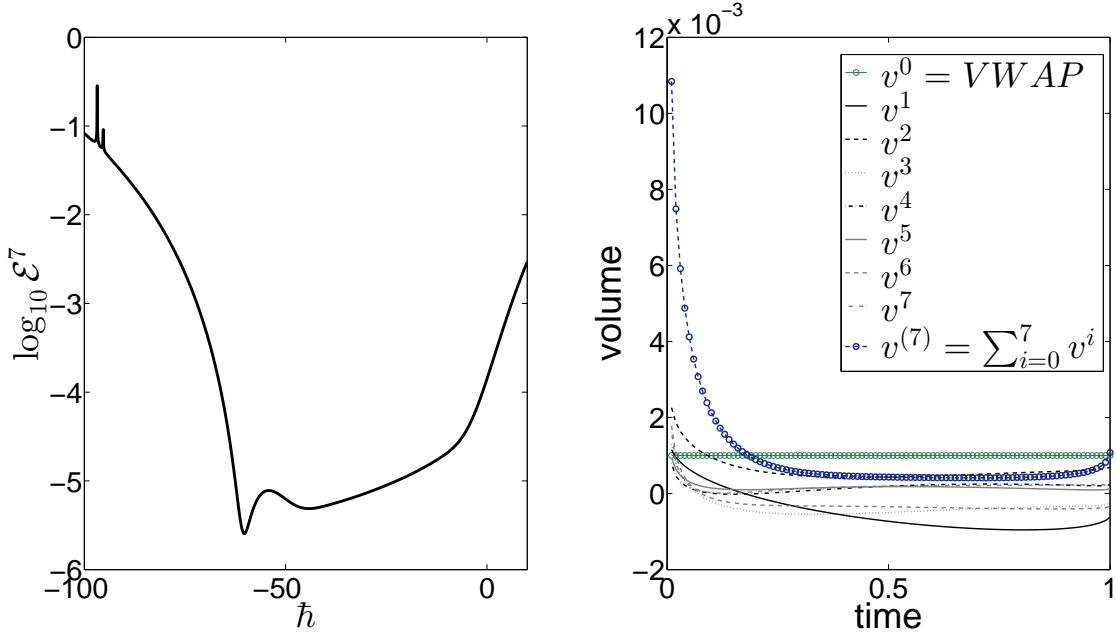} 
\vspace*{8pt}
\caption{The logarithm of the squared residual $\mathcal{E}^{7}\left(\hslash\right)$ is illustrated on the left panel, the minimum is 
         attained for $\hslash=-60.3$ where we have $\mathcal{E}^{7}=2.5\times10^{-6}$. The VWAP initial guess and the DHAM solution are reported
         on the right panel respectively by a full green line with circles and a dashed blue line with circles, are reported also the results of the seven deformation equations.}
\label{fig.1}
\end{figure}     

\begin{figure}[th]
\includegraphics[width=1.0\textwidth]{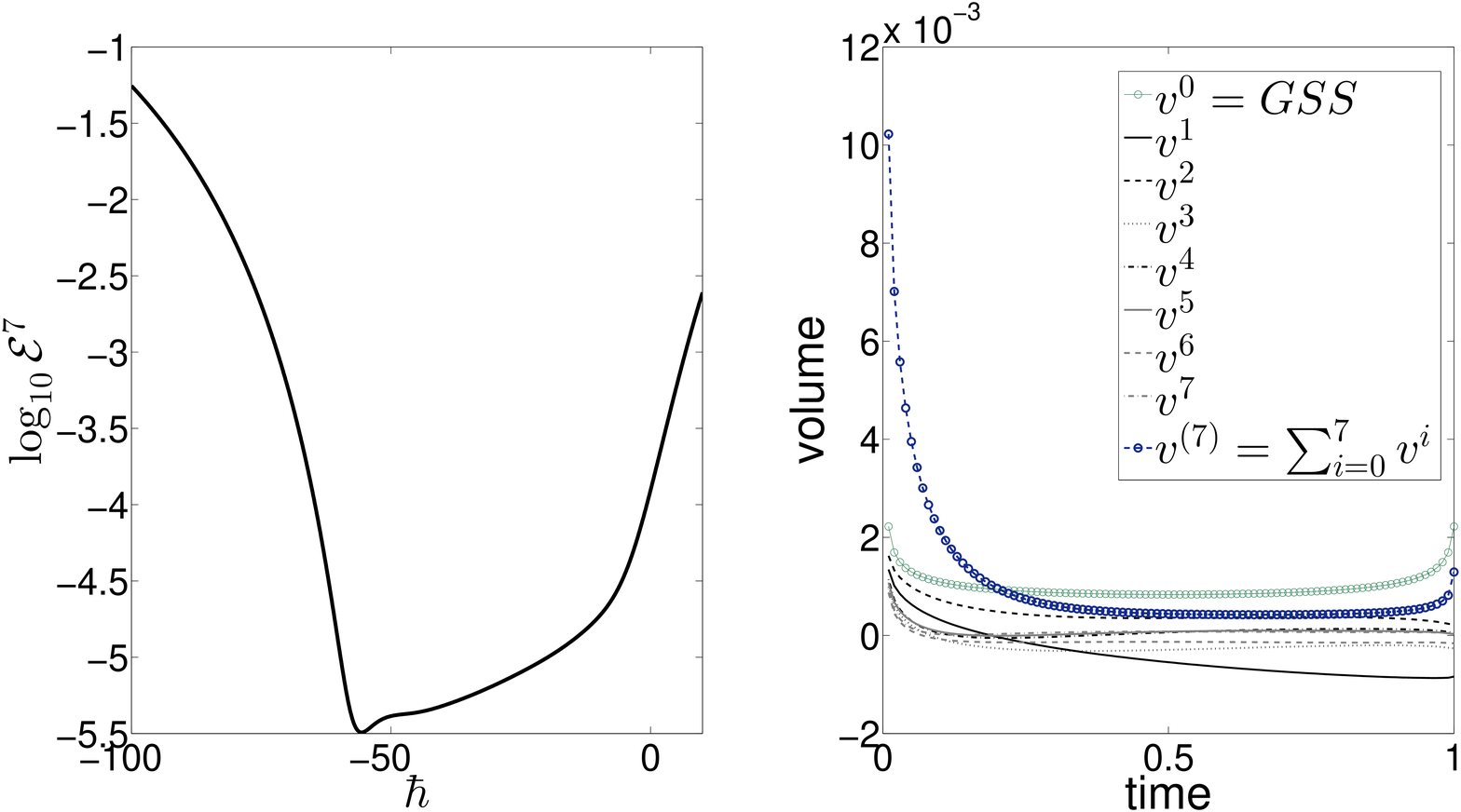} 
\vspace*{8pt}
\caption{The logarithm of the squared residual $\mathcal{E}^{7}\left(\hslash\right)$ is illustrated on the left panel, the minimum is 
         attained for $\hslash=-55.7$ where we have $\mathcal{E}^{7}=3.2\times10^{-6}$. The GSS initial guess and the DHAM solution are reported
         on the right panel respectively by a full green line with circles and a dashed blue line with circles, are reported also the results of the seven deformation equations.}
\label{fig.2}
\end{figure}

The results for a VWAP initial condition are illustrated in Figure \ref{fig.1}. The left panel shows that the  minimum of squared residual of the $7$-th order iteration is attained at $\hslash=-60.3$.  The results for a GSS initial condition are illustrated in Figure \ref{fig.2}, where the minimum of the squared residual is attained at $\hslash=-55.7$ (left panel). 
The right panel of both figures shows the trading profile of the different approximating terms $v^i$, ($i=1,...,7)$, as well as of the $7$-th order approximated solution $v^{(7)}$ obtained by summing the terms. It is worth noticing that higher order terms become smaller and smaller. In both cases the DHAM approximated solution is asymmetric with respect to time reversal; it is optimal to trade faster at the beginning of the trading period and slower at the end. This is consistent with our earlier findings in Section \ref{Perturbative approach} for small deviations from linear impact. Finally, it is worth noticing that the two different initial conditions lead to very similar approximated solutions. In fact, if we exclude the last discretization point ($t_{100}=T$), the maximum value of the relative difference between the two profiles is around $8\%$, but the relative difference of the expected liquidation cost, given by equation \eqref{eq:3.35}, is only $0.5\%$.

We now discuss the advantage of a DHAM execution with respect to VWAP\footnote[2]{The cost of a VWAP execution is $C_{VWAP}=X\,f\left(X\right)T^{\left( 1-\gamma \right)}/\left(\left(1-\gamma\right)\left(2-\gamma\right)\right)$.} and GSS executions in terms of expected liquidation costs. 
The expected liquidation costs of the three strategies are reported in Table \ref{Table.0}, where we consider $\gamma=0.45$ and $0.5$ and $\delta\in [1/2, 1]$. 
As expected, close to the linear case ($\delta=1$) the difference in cost between the DHAM and the GSS is negligible\footnote[3]{In the linear case, i.e. $\delta=1$, is possible to find a DHAM strategy with a cost lower than that
of GSS strategy. This is only a numerical artifact, because here we implement a discretized version of the GSS strategy defined by \eqref{lineargss}.
We compute the cost of a constant piece-wise approximation of the GSS. This approximation is then used as the initial guess of the DHAM solution.
In the linear case, when the value of $N$ increases the costs of the approximated GSS and of DHAM strategies become equal.}. 
Interestingly, the VWAP strategy, which is clearly different from the GSS, has a cost essentially equivalent. Thus in the linear case, the advantage of sophisticated execution strategies relative to a straightforward VWAP is very small. In contrast, as we move toward the strongly nonlinear regime, {\em i.e.} $\delta\approx0.5$, the DHAM solution achieves the smallest cost, while the worst strategy is the VWAP.   As can be observed from Table \ref{Table.0}, the improvement of DHAM with respect to GSS is much larger than the improvement of GSS with respect to VWAP. For example, for $\delta=\gamma=0.5$, the DHAM has a cost 20\% smaller than the GSS, while the latter has a cost which is only 1\% smaller than the VWAP. 

\begin{table}
\small 
\centering
\begin{tabular}{ |l|l|l|l||l|l|l|}
\hline
           & VWAP & GSS & DHAM & VWAP & GSS & DHAM \\
\hline
 $\delta$  & $\gamma=0.45$ & $\gamma=0.45$ & $\gamma=0.45$ & $\gamma=0.5$ & $\gamma=0.5$ & $\gamma=0.5$ \\
\hline
 $1.0$ & $0.0117$ & ${\bf 0.0116}$  & ${\bf 0.0116}$ & $0.0133$ &$0.0132$ &  ${\bf 0.0131}$ \\
\hline
 $0.95$ & $0.0132$ & ${\bf 0.0130}$& ${\bf 0.0130}$& $0.0150$ &  ${\bf 0.0148}$ &  ${\bf 0.0148}$ \\
\hline
 $0.90$ & $0.0148$  & $0.0146$ & ${\bf 0.0143}$ & $0.0168$ & $0.0166$  & ${\bf 0.0164}$ \\
\hline
 $0.85$ & $0.0166$ & $0.0164$ & ${\bf 0.0162}$  & $0.0188$  & $0.0186$ & ${\bf 0.0185}$ \\
\hline
 $0.80$ & $0.0186$  & $0.0184$ & ${\bf 0.0179}$ & $0.0211$ & $0.0209$  & ${\bf 0.0204}$ \\
\hline
 $0.75$ & $0.0209$  & $0.0206$ & ${\bf 0.0198}$ & $0.0237$ & $0.0234$  & ${\bf 0.0227}$ \\
\hline
 $0.70$ & $0.0234$  & $0.0231$ & ${\bf 0.0218}$ & $0.0266$& $0.0263$  & ${\bf 0.0249}$ \\
\hline
 $0.65$ & $0.0263$  & $0.0260$ & ${\bf 0.0235}$ & $0.0298$& $0.0295$  & ${\bf 0.0274}$ \\
\hline
 $0.60$ & $0.0295$  & $0.0291$ & ${\bf 0.0251}$ & $0.0335$ & $0.0331$  & ${\bf 0.0297}$ \\
\hline
 $0.55$ & $0.0331$  & $0.0327$ & ${\bf 0.0275}$ & $0.0376$& $0.0372$  & ${\bf 0.0323}$ \\
\hline
 $0.50$ &           &          &   & $0.0422$ &$0.0417$ & ${\bf 0.0347}$ \\
\hline
\end{tabular}
\caption{Costs for three different strategies, VWAP, GSS, and DHAM, in the no-dynamic-arbitrage region for $\gamma=0.45,\,0.5$. 
         The numbers in boldface indicate strategies achieving the lowest expected cost. 
         The difference between expected costs increases with the degree of non-linearity.
         In each case we use a GSS initial guess to obtain the DHAM solution.}
\label{Table.0}
\end{table}
    
    To summarize, solutions using the DHAM approach display the same time asymmetry that we found earlier using a simple perturbative approach. More importantly, the DHAM approach allows us to compute the optimal strategy in the strong nonlinear regime, achieving expected execution costs which are significantly smaller than those obtained with VWAP or GSS solutions. Once again however, the DHAM solution does not necessarily correspond to a global minimum of the cost functional \eqref{eq:2.2}, because it is a continuous deformation of a continuous and positive initial guess. In order to find potentially lower cost strategies, in Section \ref{sec:GeneralPiecewise}, we will tackle the problem of direct numerical minimization of the cost function.

\section{Numerical optimization}
\label{sec:GeneralPiecewise} 

In this section we study the cost optimization problem in the class of general piecewise constant strategies. In this case \textit{general} means that these strategies are not necessarily deformations of continuous functions of time, such as those that can be obtained with the homotopy method. 
For this purpose we discretize the interval $[0,T]$ exactly as in Section \ref{sec:DiscreteHomotopy} and minimize the cost function  \eqref{eq:3.35} with respect to the trading rates $v_i$ subject to the constraint that the total quantity traded should be $X$. Recall that when the instantaneous impact function is linear, {\em i.e.} $f\left(v\right)\propto v$, the cost reduces to a $N$-dimensional quadratic form.  In the nonlinear case, the numerical minimization involves finding the extrema of a complicated nonlinear function of $N$ variables. We begin in Section \ref{sec:toy} by presenting a motivating toy example that demonstrates how a nonlinear instantaneous market impact function $f$ can lead to transaction-triggered price manipulation, {\em i.e.} optimal strategies where it is optimal to sell in some subintervals during a buy program. In the following subsection we consider the general case. 

\subsection{A motivating example}
\label{sec:toy}

Here we show that with a nonlinear market impact function, the convexity condition on the decay kernel $G\left(\tau\right)$ is not sufficient to ensure the absence of transaction-triggered price manipulations for all the values $\gamma+\delta\geq1$, {\em i.e.} the no-dynamic-arbitrage parametric zone. 
We illustrate this concept in the simplest case with $N=2$ where the execution consists of two interval VWAPs.  Setting $T=1$, the expected liquidation cost \eqref{eq:3.35} becomes
\begin{equation}
 \label{eq:4.1}
C\left[v_{1},v_{2}\right]=\frac{1}{\left(1-\gamma\right)\left(2-\gamma\right)}\left(\frac{1}{2}\right)^{2-\gamma}\left[v_{1}f\left(
v_{1}\right)+v_{2}f\left(v_{2}\right)+\left(2^{2-\gamma}-2\right)v_{2}f\left(v_{1}\right)\right].
\end{equation}

\noindent When $f(v) \propto v$, \eqref{eq:4.1} reduces to the formula for a paraboloid in two dimensions. Imposing the constraint $v_{1}+v_{2}=2X$, the problem reduces to a minimization with respect to $v_1$.

\begin{figure}[t]
\includegraphics[width=1.0\textwidth]{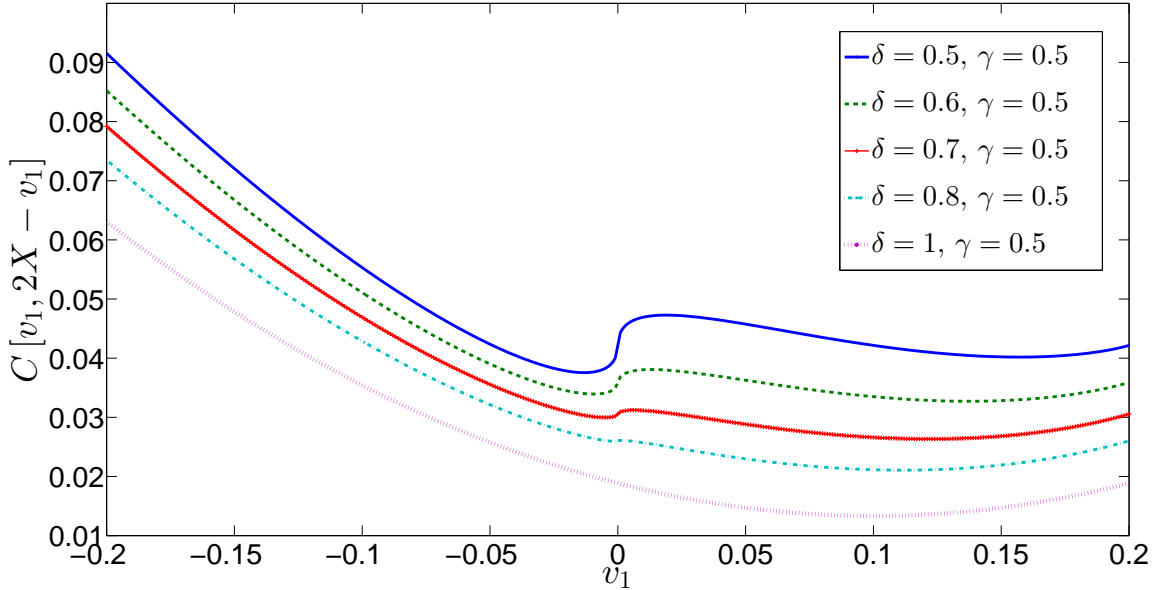}
\vspace*{8pt}
\caption{The cost function $C[v_{1},2X-v_{1}]$ for $X=0.1,\gamma=0.5$. The blue graph refers to $\delta=0.5$, green to $\delta=0.6$,
         red to $\delta=0.7$, cyan to $\delta=0.8$ and purple to $\delta=1$ where the minimum is given by $v_1=X$. In the nonlinear case
         there are two local minima.} 
\label{fig.5}
\end{figure}

In Figure \ref{fig.5}, with $\gamma=0.5$, $X=0.1$, we plot  $C\left[v_{1},v_{2}\right]$ against $v_1$ for various values of $\delta$. We observe that for $\delta<1$ the cost function has two local minima, one for a positive value of $v_1$ and one for a negative value of $v_1$. When $\delta\gtrsim 0.56$  the global minimum is the one with $v_1>0$, while for $\delta\lesssim 0.56$ the global minimum is attained for a negative value of $v_1$. It follows that if the impact function is strongly nonlinear, we can decrease the expected cost of a buy program with an intermediate sell trade, {\em i.e.} there is transaction-triggered price
manipulation. Notice that, if we impose $v_{1}\ge0$, we have a boundary solution, {\em i.e.} it is optimal not to trade in the first interval, trading the whole order in the second interval.
 
We observe this behavior for all values of the parameters near the boundary of the  no-dynamic-arbitrage region $\gamma+\delta\geq1$, {\em i.e.} in the strong nonlinear region. 
This simple example also demonstrates that in the strong nonlinear regime, in the case of a buy program, it is better to sell during the first half of the trading period and then buy during the second half. In the next section we show numerically that this effect is accentuated when $N$ is large.

\subsection{Numerical cost minimization}

In the general case we perform a non-convex constrained optimization of the cost function defined by the $N$ variables $v_i$. The linear market impact case, {\em i.e.} $\delta=1$, is a convex optimization problem, while in the non linear case the cost function is not convex and therefore we need to resort to numerical methods. 
We make use of two incomplete methods, {\em i.e.} we do not reach the global minimum
with certainty. The first method is based on the Sequential Quadratic Programming (SQP) algorithm. This is one of the most successful methods for the numerical solution of constrained nonlinear
optimization problems (NLP). It is an iterative procedure which models the NLP for a given iterate by a quadratic programming (QP) sub-problem, and 
then uses the solution to construct a new iterate. Convergence to a local minimum is then guaranteed.  We use the routines implemented in Matlab \cite{Matlab1,Matlab2,Matlab3}. 

When we search for strategies with the constraint that all trades should have the same sign, an algorithm based on derivatives, like SQP, can fail on the hyperplanes of the state space defined by $v_{i}=0$, where the derivatives of the cost function diverge.  For this case, we employ a second method based on a direct search approach, the {\em generating set search} (GSS) algorithm \cite{Kolda,Kolda2}, implemented in Matlab. This method does not require the computation of the derivatives of the cost function, because it searches directly
directions of space where the cost decreases. When the search is close to the boundary of the feasible region, the set of search directions must include directions that conform to the geometry of the boundary. As would be expected, the GSS algorithm is more computationally expensive  than the SQP algorithm. 
  
We adopt a multiple random start approach, consisting of picking random starting points on the hyperplane defined by equation \eqref{eq:3.29} (the constraint on the total quantity traded) and performing local SQP and direct search optimizations starting from these points. 
We then study the difference between the various extreme points, investigating whether they are local minima, and then among these extrema we select the solution with the smallest cost. 
Again, there is no guarantee that this corresponds to the global minimum.

\subsection{Results}

We tried different distributions for the initial points ({\em e.g.} a Dirichlet distribution on the hyperplane with various parameters), but we found qualitatively similar results in all cases. 
We illustrate here detailed results of such a procedure for $N=100$ and $\gamma=\delta=0.5$, $\gamma=0.45,\delta=0.55$ using $1000$ starting points distributed uniformly on the hyper-plane $\sum_{i=1}^Nv_i=$const. 
This numerical analysis highlights three main features of the expected cost function \eqref{eq:3.35} subject to the trading-volume constraint \eqref{eq:3.29}: (i) it has many local minima, (ii) there are many minima for which some $v_i<0$, and (iii) the optimal cost can assume negative values for a buy-program. 
We consider first the properties of the solution with the minimal cost among those found with our numerical method and then we investigate the properties of the landscape and the characterization of the suboptimal solutions.
Lastly, in Section \ref{sec:monotone}, we describe the results of optimization with the  additional constraint that $v_{i}\geq 0$ (for a buy metaorder). We show that in this case, the optimal trading strategy is \textit{sparse} when the model is strongly nonlinear, {\em i.e.} it is optimal to trade in few intense bursts.

\subsubsection{Minimal cost solution}

\begin{table}
\footnotesize
\centering
\begin{tabular}{ |l|l|l|l||l|l|l|}
\hline
           & DHAM & SQP & DIRECT & DHAM & SQP & DIRECT  \\
\hline
 $\delta$  & $\gamma=0.45$ & $\gamma=0.45$ & $\gamma=0.45$ & $\gamma=0.5$ & $\gamma=0.5$ & $\gamma=0.5$ \\
\hline
 $1.0$  & $0.0116$ & ${\bf 0.0115}$ & $0.0115$ & $0.0131$ & ${\bf 0.0131}$ & $0.0131$ \\
\hline
 $0.95$ & $0.0130$ & ${\bf 0.0128}$ & $0.0129$ & $0.0148$ & ${\bf 0.0147}$ & $0.0147$ \\
\hline
 $0.90$ & $0.0143$ & ${\bf 0.0136}$ & $0.0140$ & $0.0164$ & ${\bf 0.0158}$ & $0.0162$ \\
\hline
 $0.85$ & $0.0162$  & ${\bf 0.0139}$ & $0.0151$ & $0.0185$ & ${\bf 0.0166}$ & $0.0176$ \\
\hline
 $0.80$ & $0.0179$ & ${\bf 0.0138}$ & $0.0162$ & $0.0204$ & ${\bf 0.0170}$ & $0.0188$ \\
\hline
 $0.75$ & $0.0198$ & ${\bf 0.0132}$ & $0.0169$ & $0.0227$ & ${\bf 0.0169}$ & $0.0202$ \\
\hline
 $0.70$ & $0.0218$ & ${\bf 0.0117}$ & $0.0184$ & $0.0249$ & ${\bf 0.0163}$ & $0.0220$ \\
\hline
 $0.65$ & $0.0235$ & ${\bf 0.0092}$ & $0.0191$ & $0.0274$ & ${\bf 0.0146}$ & $0.0238$ \\
\hline
 $0.60$ & $0.0251$ & ${\bf 0.0047}$ & $0.0201$ & $0.0297$ & ${\bf 0.0120}$ & $0.0245$ \\
\hline
 $0.55$ & $0.0275$ & ${\bf -0.0029}$ & $0.0212$ & $0.0323$ & ${\bf 0.0075}$ & $0.0262$ \\
\hline
 $0.50$ &          &                 &          & $0.0347$ & ${\bf 0.0003}$ & $0.0278$ \\
\hline
\end{tabular}
\caption{Costs of three different strategies, DHAM, SQP and Direct-search in the no-dynamic-arbitrage region for $\gamma=0.45,\,0.5$. The numbers in boldface indicate strategies achieving the lowest expected cost. The difference between expected costs increases with the degree of non-linearity.
         In each case we use a GSS initial guess to obtain the DHAM solution. We use a discretization of $N=100$ subintervals and the SQP and Direct-search optimization are performed by using $1000$ starting points.}
\label{Table.3}
\end{table}

\begin{figure}[t]
\includegraphics[width=1.0\textwidth]{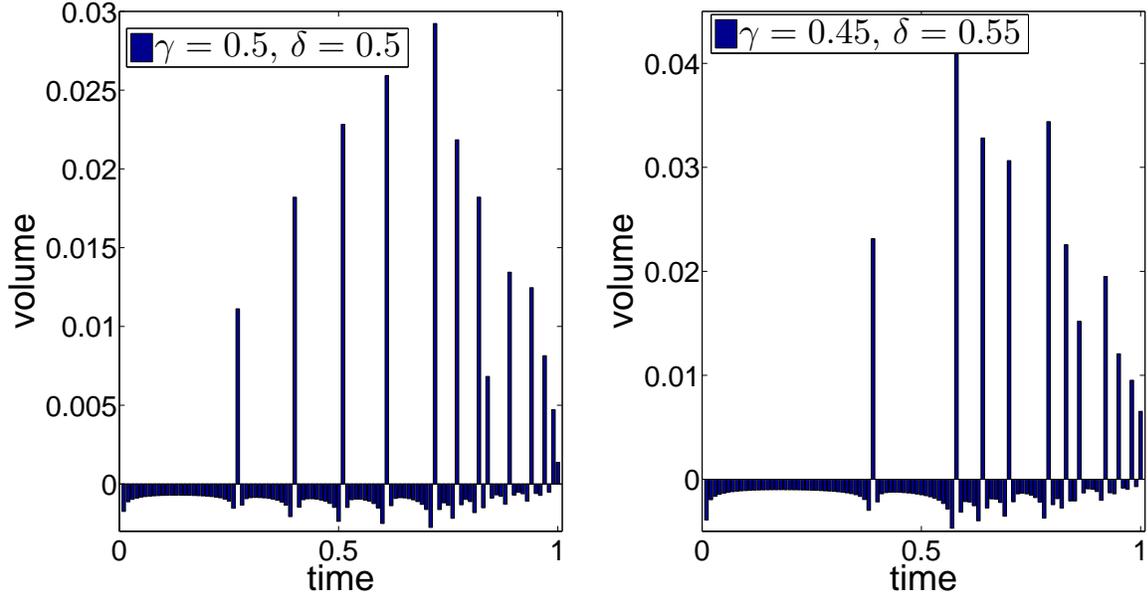}
\vspace*{8pt}
\caption{Optimal solution given by the SQP-algorithm for a buy-program where $X=0.1$, {\em i.e.} $10\%$ of a unitary market volume. We report the volume to be traded in each interval of time, {\em i.e.} $v_{i}\,T/N$.}
\label{fig.8}
\end{figure}

In Figure \ref{fig.8}, we plot the optimal trading profile for a buy program corresponding to the global minimum of the cost function using our numerical minimization procedure. We observe that the optimal quantity to trade in each subinterval varies very irregularly with time. 
The cost minimizing solution consists of a series of bursts of intense but short-lived buying, separated by long periods when it is optimal to sell slowly. Evidently, the optimal solution admits transaction-triggered manipulation, as already shown in the toy $N=2$ case analyzed above.

\begin{figure}[t]
\includegraphics[width=1.0\textwidth]{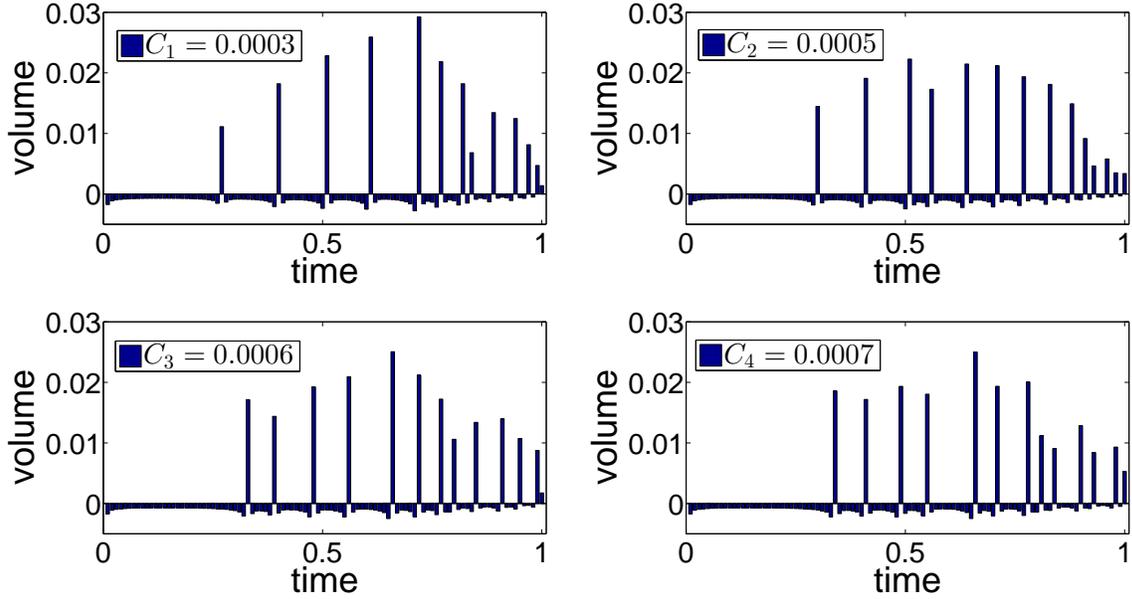}
\vspace*{8pt}
\caption{The four lowest cost solutions given by the SQP-algorithm for a buy-program where $X=0.1$, {\em i.e.} $10\%$ of a unitary market volume for $\gamma=0.5,\,\delta=0.5$. We report the volume to be traded in each interval of time, {\em i.e.} $v_{i}\,T/N$.
The costs are reported in the insets.}
\label{fig.8.0}
\end{figure}

Strategies that are close-to-optimal are qualitatively similar, but the positions of their spikes can be very different. As an example, in Figure \ref{fig.8.0}, we plot the four lowest cost solutions when $\gamma=0.5$ and $\delta=0.5$. All are characterized by a few intense positive spikes, separated by periods of slow selling. But these solutions differ in the positions of the bursts of buying. To quantify the differences between these solutions, we compute the Euclidian distance between them, including a simple VWAP strategy for reference. The resulting distance matrix is

$$
D=
\begin{pmatrix}
0      & 0.0780 & 0.0753 & 0.0890 & 0.0617 \\
0.0780 & 0      & 0.0757 & 0.0757 & 0.0582 \\
0.0753 & 0.0757 & 0      & 0.0799 & 0.0581 \\
0.0890 & 0.0757 & 0.0799 & 0      & 0.0576 \\
0.0617 & 0.0582 & 0.0581 & 0.0576 & 0  
\end{pmatrix},
$$    
where the fifth row and column refer to the VWAP. We notice that all distances are quite similar.  Strikingly however, the distance between any two SQP strategies in Figure \ref{fig.8.0} is greater than the distance of either of these to a simple VWAP; there is typically a multiplicity of strategies with bursts at different times, but with similar expected costs.

\begin{figure}[t]
\begin{center}
\includegraphics[width=0.6\textwidth]{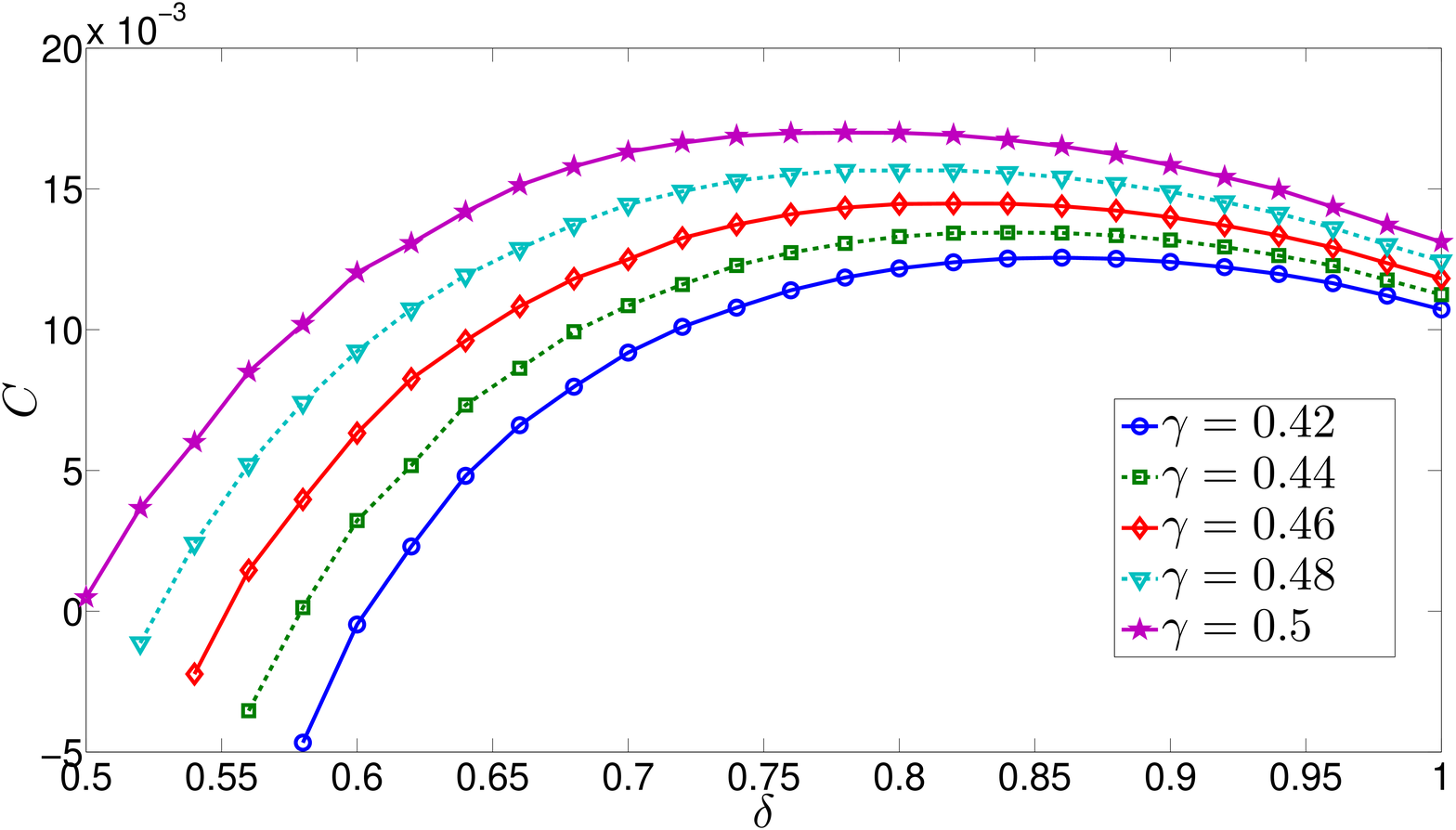}\\
\includegraphics[width=0.6\textwidth]{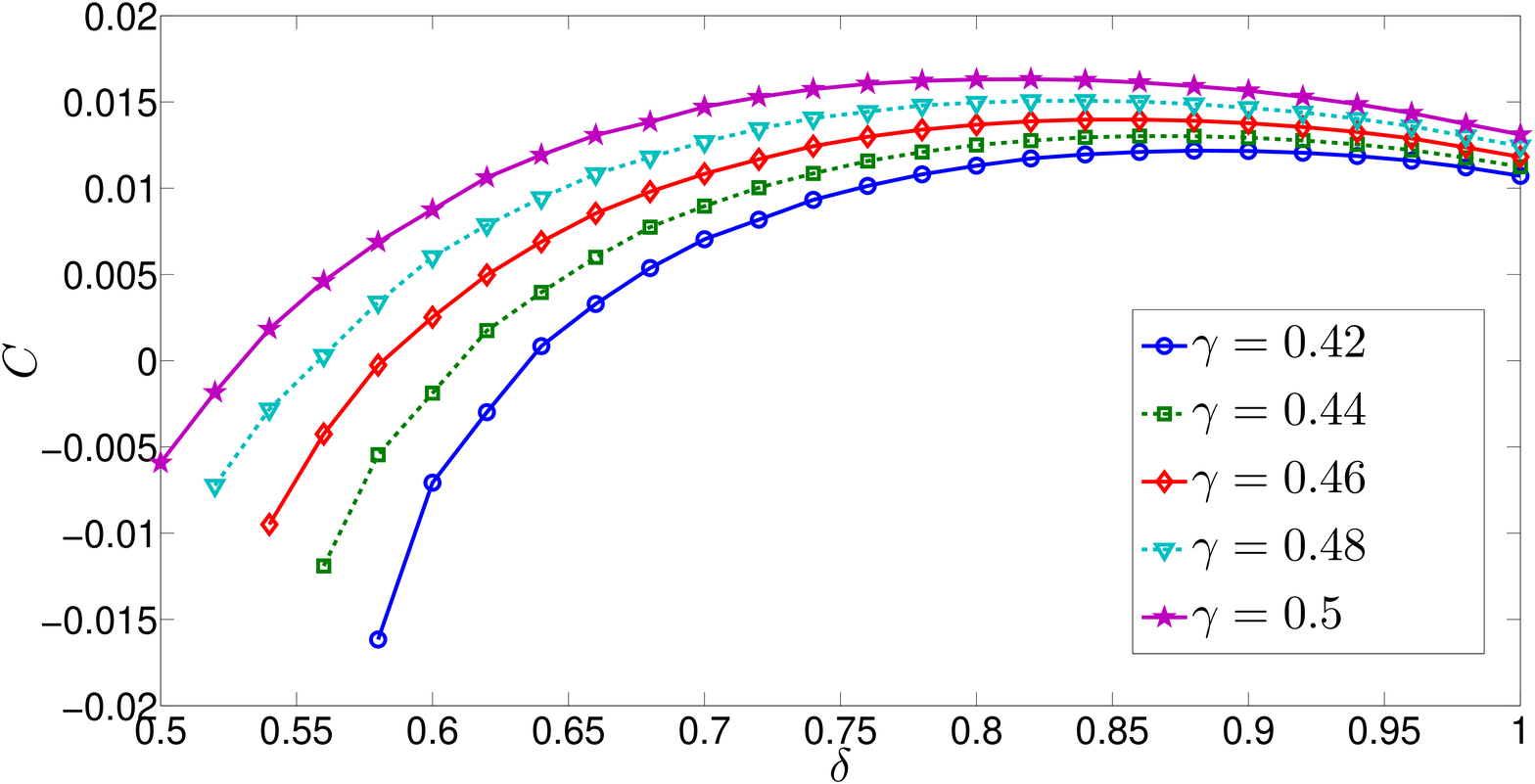}
\end{center}
\vspace*{8pt}
\caption{Optimal cost for the nonlinear transient impact model obtained with the SQP minimization. We consider a buy-program where $X=0.1$, {\em i.e.} $10\%$ of a unitary market volume and a discretization of $N=100$ (top) and $N=150$  (bottom) subintervals. 
         We use $1000$ starting points for each optimization. We consider only parameter in the no-dynamic-arbitrage zone.
         Holding $\gamma$ fixed, the cost relative to the global minima is not a monotonic function of $\delta$.} 
\label{fig.8.1}
\end{figure}

In Table \ref{Table.3} we report the costs for these various cost-minimizing solutions.
We observe that the expected costs of SQP solutions are close to zero and even negative in the case $\gamma=0.45,\,\delta=0.55$. To further explore this undesirable behavior, in Figure \ref{fig.8.1} we plot the expected minimal cost for different values of $\gamma$ as a function of $\delta$ and $N=100$ (top panel) and $N=150$ (bottom panel). First, we observe that costs are not a monotonic function of $\delta$. Second, when $N=100$ for $\gamma<0.5$ we find a regime of $\delta$ for which the expected cost is negative. This effect becomes stronger when we increase the discretization. In fact, when $N=150$ even for $\gamma=0.5$ it is possible to find values of $\delta$ for which the minimal cost is negative. 


A negative expected cost is a signature of the possibility of price manipulation. Indeed if market impact is entirely transient, with no permanent component, as in the model under consideration, it would then be possible to devise a round trip strategy with positive (expected) revenue, {\em i.e.} a wonderful money machine. We cannot then escape the conclusion that a nonlinear transient impact model \eqref{eq:2.1} with $f(v)=v^\delta, \, \delta <1$ is mis-specified since it allows arbitrage opportunities.

\subsubsection{Characterization of the cost landscape}

We now investigate the structure of the cost landscape to be minimized and the properties of the extrema.  We find a very large number of distinct extremal points. 
Jarque-Bera and chi-square goodness-of-fit tests cannot reject at the $5\%$ significance level the Gaussian hypothesis for the distribution of cost of the local minima in the two data sets\footnote[4]{The 
Jarque-Bera test gives a $p$-value$=0.22$ for the case $\gamma=\delta=0.5$ and a $p$-value$=0.49$ for the case $\gamma=0.45,\,\delta=0.55$.
The chi-square goodness-of-fit test gives a $p$-value$=0.36$ for the case $\gamma=\delta=0.5$ and a $p$-value$=0.93$ for the case $\gamma=0.45,\,\delta=0.55$.}. Thus we find a large number of extremal points which are quite similar in terms of expected cost. As expected the standard deviation of the cost of the local minima decreases as the market impact function becomes more linear. Similarly, the Euclidean distance between the local minima decreases when the model is closer to linear.

In general, the presence of many minima could reflect numerical error or a landscape that truly features multiple minima. The first case occurs frequently when the landscape is {\it sloppy} \cite{Brown,Waterfall}, {\em i.e.} when there are directions of the  space of variables along which the cost is substantially unchanged. A well known example of such a behavior is given by the Rosenbrock function, in which the global minimum is located inside a long narrow and parabolic shaped valley. Sloppiness can imply difficulties in  finding the global minimum, because there is a manifold where the cost function is almost flat. Alternatively, the landscape may be {\it rugged}, with many local minima separated by local peaks. Rugged landscapes have attracted attention in physics \cite{Weinberger}, evolutionary biology \cite{Weinberger2}, and computer science. An example of a rugged landscape is Kauffman's $N$-$k$ fitness landscape \cite{Weinberger}. The distribution of locally minimal energies of its  spin-glass Hamiltonian is described by a normal distribution. In order to discriminate between these two alternatives and to characterize the cost landscape we performed two types of analysis.

First, we apply the second order condition to test which of the found extrema are actually minima and not, for example, saddle points. We perform a second derivative test on the local minima found by the SQP algorithm. In the case of constrained optimization the second-order sufficient condition for a minimum can be expressed in a determinant form of the bordered Hessian \cite{Chiang,Magnus}. 
We report the details of this analysis for the concave impact case
when we choose a discretization $N=100$. We have repeated this analysis for the concave-convex impact (see below) case finding qualitatively the same results. Our optimization program starts from $1,000$ initial points, so we compute the above test on the final points where the SQP algorithm converges.  For any value of the parameters $\delta$ and $\gamma$ in the no-arbitrage region, more than 95\% of the extremal points are actual minima. The remaining points are saddle points.


Second, we directly test the hypothesis that the landscape is sloppy, by using the eigenvalues and eigenvectors of the bordered Hessian of the lagrangian landscape function to identify its \textit{stiff} and \textit{sloppy} directions. One can study the sensitivity of the landscape function to changes by the eigenvalue spectra of the Hessian computed at a local minimum. Sloppy models are described by a constant logarithmic density for eigenvalues over six or more order of magnitude \cite{Brown,Waterfall}. The sensitivity of the landscape to changes is given by the square root of the eigenvalue. For sloppy models this means that one should move along the sloppiest eigen-direction a thousand times more than along the stiffest eigen-direction in order to change the function by the same amount.  We have computed eigenvalue spectra of the bordered Hessian of the lagrangian function in the case of a concave market impact function and we found that the spectra are not compatible with a sloppy landscape. The bulk of the eigenvalues have similar small values of the order of $10^{-2}$. This implies that the region near a local minimum is not flat. 

In summary, the above analyses indicate that the vast majority of extremal points are actual minima and the region around a local minimum is not flat shaped in some direction. This suggests that the cost landscape under concave instantaneous market impact 
is rugged rather than sloppy. This situation is reminiscent of the search for the global minimum of the free-energy force field describing a protein \cite{Verma}. The similarity with our problem is given by the structure of the free-energy, described by the sum of many Lennard-Jones potentials, {\em i.e.} a non-convex and non-periodic function. The great number of local minima is a consequence of summing non-convex functions and not a consequence of the presence of periodic functions. Similarly, in the concave impact case the cost is the sum of functions like the one described in Figure \ref{fig.5} for our $2$-D toy model; such functions are non-convex and non-periodic.

\subsubsection{Monotone strategies}\label{sec:monotone}

In this section we consider monotone strategies, {\em i.e.} strategies where the position $X(t)$ is a non decreasing (non increasing ) function of time for buy (sell) metaorders. We thus impose the absence of transaction-triggered manipulation. From the point of view of the numerical cost minimization, we impose the additional constraints  $v_{i}\geq 0$, $\forall i$ (for a buy program). This additional constraint is analogous to the no-short-selling constraint  in portfolio optimization \cite{Brodie}.

\begin{figure}[t]
\includegraphics[width=1.0\textwidth]{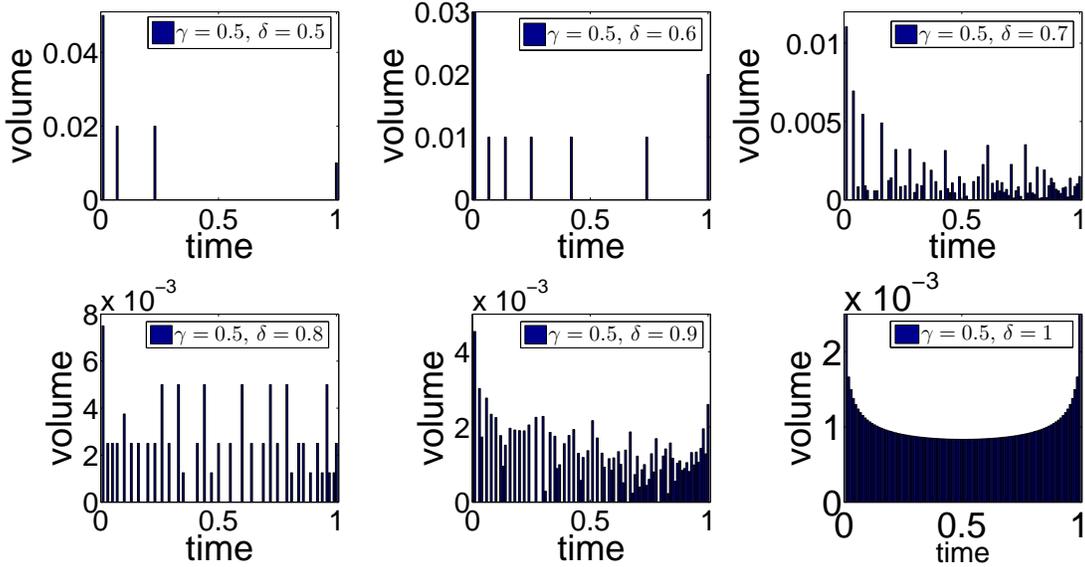}
\vspace*{8pt}
\caption{Optimal solutions given by the direct-search algorithm for a monotone buy-program where $X=0.1$, {\em i.e.} $10\%$ of a unitary market volume for $\gamma=0.5$. We report the volume to be traded in each interval of time, {\em i.e.} $v_{i}\,T/N$.}
\label{fig.8.2}
\end{figure}

We use a direct search method, specifically the generating set search algorithm \cite{Kolda2}. Obviously, the cost is positive in this case; we report the costs of optimal strategies in Table \ref{Table.3}. We find that the expected cost increases as the instantaneous market impact function becomes more nonlinear ({\em i.e.} as $\delta$ decreases).  Moreover, expected costs are higher than that found by the SQP algorithm, but still significantly lower than expected of the strategies obtained using the DHAM approach. The structure of the optimal solutions is given in Figure \ref{fig.8.2} for the case $\gamma=0.5$; results for $\gamma=0.45$ are similar. The main feature that we observe is that as $\delta$ decreases, so that the market impact function becomes more nonlinear, the optimal trading profile becomes more and more \textit{sparse}. The optimal strategy then consists of a  few bursts of buying interspersed with long periods of no-trading. The geometrical interpretation of this result is that the solution lies on
the boundary of a $(N-1)$-simplex. This is consistent with our unconstrained SQP results, where optimal strategies lie just beyond the edge of the simplex. The direct search simply stops at the edge of the feasible region described by the simplex. Finally notice that under the monotonicity constraint, it is optimal to start trading in the first interval, whilst without this constraint, it is optimal to slowly push down the price by selling before trading the first burst.  

\section{Regularizing the solution}
\label{regularize}
In this section, we show that the nonlinear impact model \eqref{eq:2.2} may be regularized using two different approaches, both of them reflecting important features of the market that we have so far neglected.  In Section \ref{sec:spread}, we add a spread cost to the model in order 
to penalize wrong-way trading that may give rise to negative execution costs. This is equivalent to an $L_1$ or LASSO regularization.
In Section \ref{sec:concave-convex} we modify the shape of the instantaneous market impact function $f$ for high trading rates; the resulting market impact function is then concave-convex.

\subsection{Adding a spread cost}
\label{sec:spread} 
In this section we add a spread cost to the model of equation \eqref{eq:2.1} to penalize wrong-way trading.  
Equation \eqref{eq:2.1} could be regarded as describing the evolution of the mid-price.  
When a market order is executed\footnote[5]{Even if execution is with limit orders, one finds in practice that due to adverse selection, there is an extra cost of a portion of the bid-ask spread.}, there is an extra cost of half of the bid-ask spread $2\delta_S$, and the trading price is given by
\begin{equation}
\label{eq:5.1}
S\left(t\right)=S\left(0\right)+\int_{0}^{t} f\left( \dot x\left(t\right)\right)G\left(t-s\right)\,ds+\int_{0}^{t} \sigma \,dW(s)+\delta_{S}\,\int_{0}^{t} \delta\left(s-t\right)sign \left(\dot x\left(s\right)\right)\,ds,
\end{equation}
The spread term is a temporary impact term that can be described by a $\delta$-impact function; it affects only the price at which the order is executed and does not affect the market price.  This term can represent also any cost or fee proportional to the absolute volume executed. 
The expected execution cost is then given by
\begin{equation}
\label{eq:5.2}
C\left[\Pi\right]=C_{G}\left[\Pi\right]+C_{S}\left[\Pi\right]=\int_{0}^{T}\dot x\left(t\right)\,\int_{0}^{t}f\left(\dot x\left(s\right)\right)G\left(t-s\right)ds\,dt+\delta_{S}\int_{0}^{T} |\dot x\left(t\right)|dt.
\end{equation}
The second term thus penalizes any strategy that consists of both buy and sell trades. This term is minimized, {\em i.e.} $C_{S}\left[\Pi\right]=\delta_{S}\,X$, for 
strategies where all trades have the same sign. This penalty is a form of $L_1$ or LASSO regularization widely used in computer science and used by Brodie et al. \cite{Brodie} to penalize short positions
in Markowitz portfolio optimization. Busseti and Lillo \cite{Busseti} have calibrated optimal execution
strategies on real data by regularizing them with a spread cost of this form.

\begin{figure}[t]
\begin{center}
\includegraphics[width=0.9\textwidth]{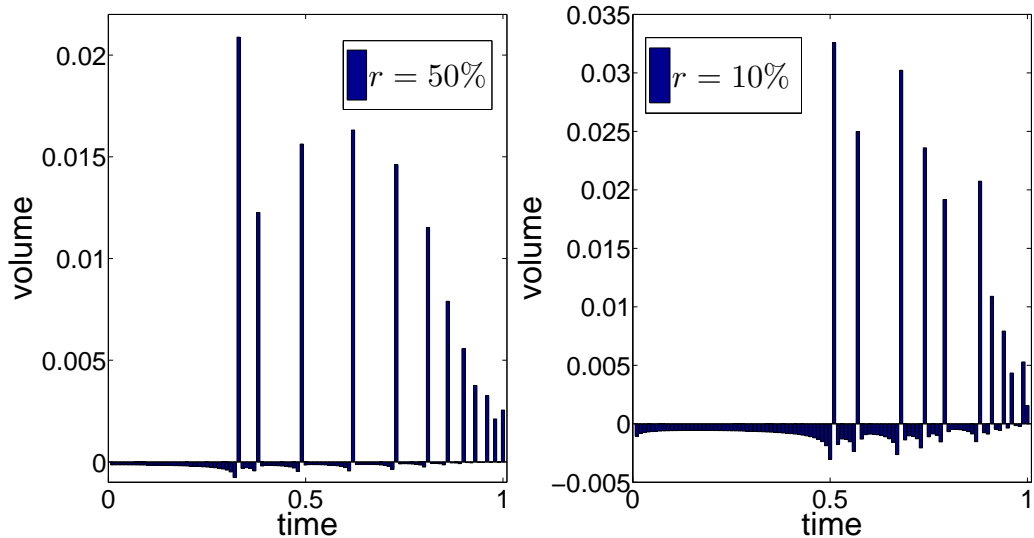}
\end{center}
\vspace*{8pt}
\caption{Optimal solution given by the SQP algorithm for a buy-program where $X=0.1$, {\em i.e.} $10\%$ of market volume, in presence of a spread cost. 
         We report the volume to be traded in each interval of time, {\em i.e.} $v_{i}\,T/N$ for the case $\gamma=0.45,\delta=0.55$.
         On the left is the case of high spread cost, {\em i.e.} $r=50\%$, on the right is  the case of low spread cost, {\em i.e.} $r=10\%$.
         The expected execution cost is $C_{SQP}=0.026$ for $r=50\%$ and $C_{SQP}=5.9 \times 10^{-3}$ for $r=10\%$.}
\label{fig.9}
\end{figure}

In order to parameterize the relative cost of spread and impact, we define the dimensionless  quantity $r=C_{S}\left[VWAP\right]/C_{G}\left[VWAP\right]$, which is the ratio between the spread cost and the impact cost
for a VWAP execution according to the  model of equation \eqref{eq:5.1}. This quantity is given by 
\begin{equation}
\delta_{S}=r\,f\left(X\right)\frac{T^{1-\gamma}}{\left(1-\gamma\right)\left(2-\gamma\right)}
\end{equation}
By choosing a value of $r$ we set the value of $\delta_{S}$. We consider
the case of small spread cost, {\em i.e.} $r=10\%$, and the case of high spread cost, {\em i.e.} $r=50\%$ and perform a numerical optimization of the discretized cost function
\begin{equation}
\label{eq:5.3}
C\left[v\right]=\sum_{i=1}^{N}\,\sum_{j=1}^{N} v_{i}\,f\left(v_{j}\right) A_{ij}+\delta_{S}\,\frac{T}{N}\,\sum_{i=1}^{N}|v_{i}|,
\end{equation}
with the same parameters as in Section \ref{sec:GeneralPiecewise}, $N=100,\,X=0.1$, for the case $\gamma=0.45$, $\delta=0.55$. Figure \ref{fig.9} plots the strategies corresponding to the global minimum in the numerical optimization in high spread cost and low spread cost cases respectively.

Despite the fact that transaction-triggered price manipulation is still observed, in both cases the liquidation cost is now positive. In fact, we find $C_{SQP}=0.026$ for $r=50\%$ and $C_{SQP}=5.9 \times 10^{-3}$ for $r=10\%$. SQP strategies still outperform VWAP, GSS and DHAM strategies.  


Finally, the effect of a high spread cost is a substantial reduction in negative trading volume similar to what is observed
by Brodie et al. \cite{Brodie} for Markowitz portfolio weights. In that case a penalization term proportional to the sum of absolute values of weights leads to optimal solutions where the resulting portfolio is sparse
with few assets and no-short positions. Similarly, in a nonlinear market impact model with a high spread-cost,
optimal solutions for a buy-program appears to be characterized by a few bursts of trading-activity separated by intervals of time in which we have a very weak negative trading activity. 
Moreover, a high spread cost leads to liquidation costs similar to those corresponding to monotone strategies. 

\subsection{Concave-convex impact}
\label{sec:concave-convex}
\begin{figure}[t]
\begin{center}
\includegraphics[width=0.9\textwidth]{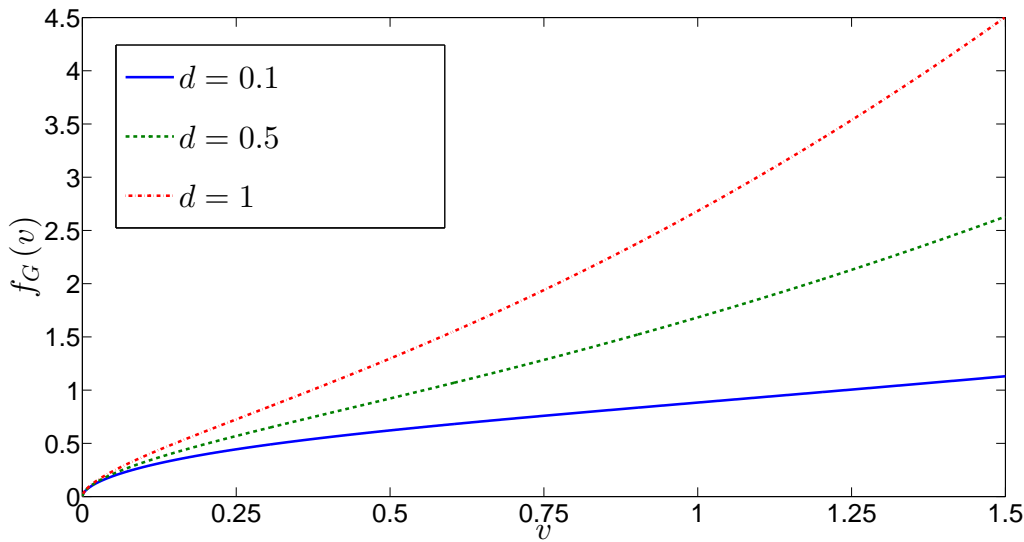}
\end{center}
\vspace*{8pt}
\caption{Concave-convex impact function for values of parameters: $c=1,\,\delta=0.55,\,X_M=1,\,T=1$.} 
\label{fig.10.0}
\end{figure}

Up to now, we have assumed that it is possible to trade arbitrarily fast whilst preserving the same functional form of the impact $f(v)$. 
Although this simplifies the mathematical treatment of the problem, it is unrealistic since we cannot trade at an arbitrarily high rate in practice. 
At some point the rate of trading is so high that one trades deeply into the order book, where less liquidity is available, and at this point one can expect that  $f(\cdot)$ becomes  convex. 
This motivates us to regularize the transient impact model by postulating a concave-convex impact function that might both penalize excessively high trading rates and potentially eliminate the problem of negative expected liquidation costs. 
Specifically, we consider the following form for the impact function:
\begin{equation}
 \label{eq:6.1}
f_{G}\left(v\right)=c\,\, sign \left(v\right)\left \{ \left(\frac{|v|}{|v|+V}\right)^{\delta}+d\frac{|v|\left(|v|+V\right)}{V^2}\right \}, 
\end{equation}
where $V=X_{M}/T$ is the market volume $X_M$ per unit time and $c$ and $d$ are positive constants. Figure \ref{fig.10.0} shows the form of this impact function. The constant $d$ is a measure of the magnitude of the convex term with respect to the concave term, which is of order one. If $d\ll1$ we recover a concave impact function. 
The parameter $d$ sets directly also the value $v^{*}$ where the function's convexity changes, where $f_{G}''\left(v^{*}\right)=0$, see second column of Table \ref{Table.6}. One could think of the rate $v^{*}$ as approximating the maximum rate at which it is reasonable to trade. 

\begin{figure}[t]
\begin{center}
\includegraphics[width=0.9\textwidth]{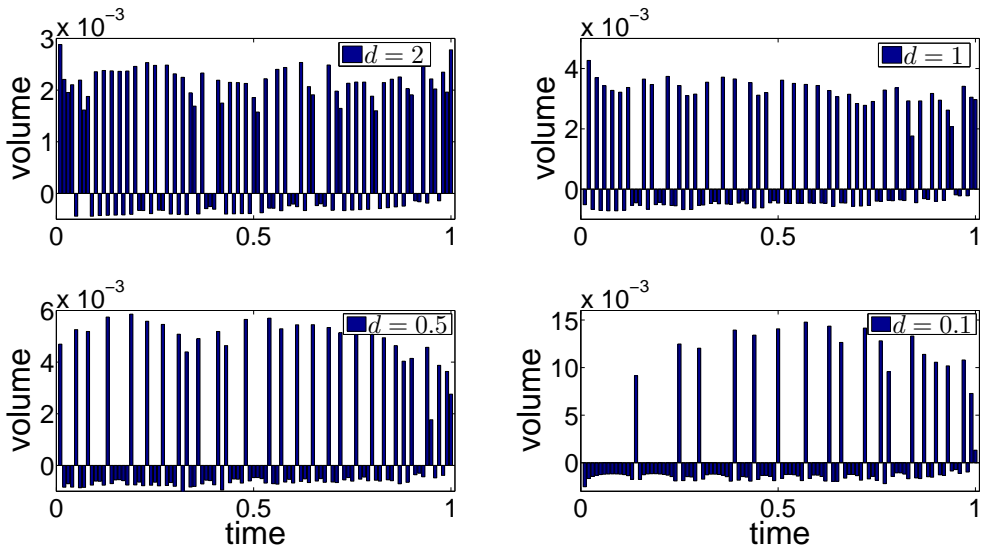}
\end{center}
\vspace*{8pt}
\caption{Optimal solution given by the SQP-algorithm for a buy-program where $X=0.1$, {\em i.e.} $10\%$ of market volume, 
         in presence of a concave-convex impact. 
         We report the volume to be traded in each interval of time, {\em i.e.} $v_{i}\,T/N$ for the case $\gamma=0.45,\delta=0.55,N=100$.
         When the magnitude of the convex term increases, {\em i.e.} as $d$ increases, trading rates decrease. 
         Expected execution costs are given in Table \ref{Table.6}.}
\label{fig.10}
\end{figure}

 We illustrate the results of numerical minimization of expected cost in the case $\gamma=0.45,\delta=0.55,N=100$ for four different values of $d$: $0.1,\,0.5,\,1,\,2$, using $1000$ starting points for each optimization.
In Fig. \ref{fig.10}, we plot the optimal strategies in each of these four cases. We observe that an increase in the magnitude of the convex term causes a decrease in maximum trading rates and an increase in the number of periods in which buying is optimal. 
The convex part of the impact acts like a barrier for high trading rates. In Table \ref{Table.6} we compare the costs of the SQP strategies with respect with the cost of the corresponding VWAP strategy  
\footnote[6]{For the concave-convex impact function the cost of VWAP is 
$C_{VWAP}=\frac{c\, X T^{\left( 1-\gamma \right)}}{\left(1-\gamma\right)\left(2-\gamma\right)}\left \{\left(\frac{X}{X_{M}+X}\right)^{\delta}+d\left(\frac{X\left(X+X_{M}\right)}{\left( X_{M} \right)^{2}}\right) \right \}.$}.
As the magnitude of the convex term increases we no longer find negative expected liquidation costs; at least in simulation,  the presence of the convex part of the impact function is able to regularize the nonlinear transient impact model. It is worth noticing that the value of $v^{*}$ is close to the mean trading rate in each case.  Thus by knowing the impact function, one has a rough estimate of the optimal trading speed. 

\begin{figure}[th]
\begin{center}
\includegraphics[width=0.5\textwidth]{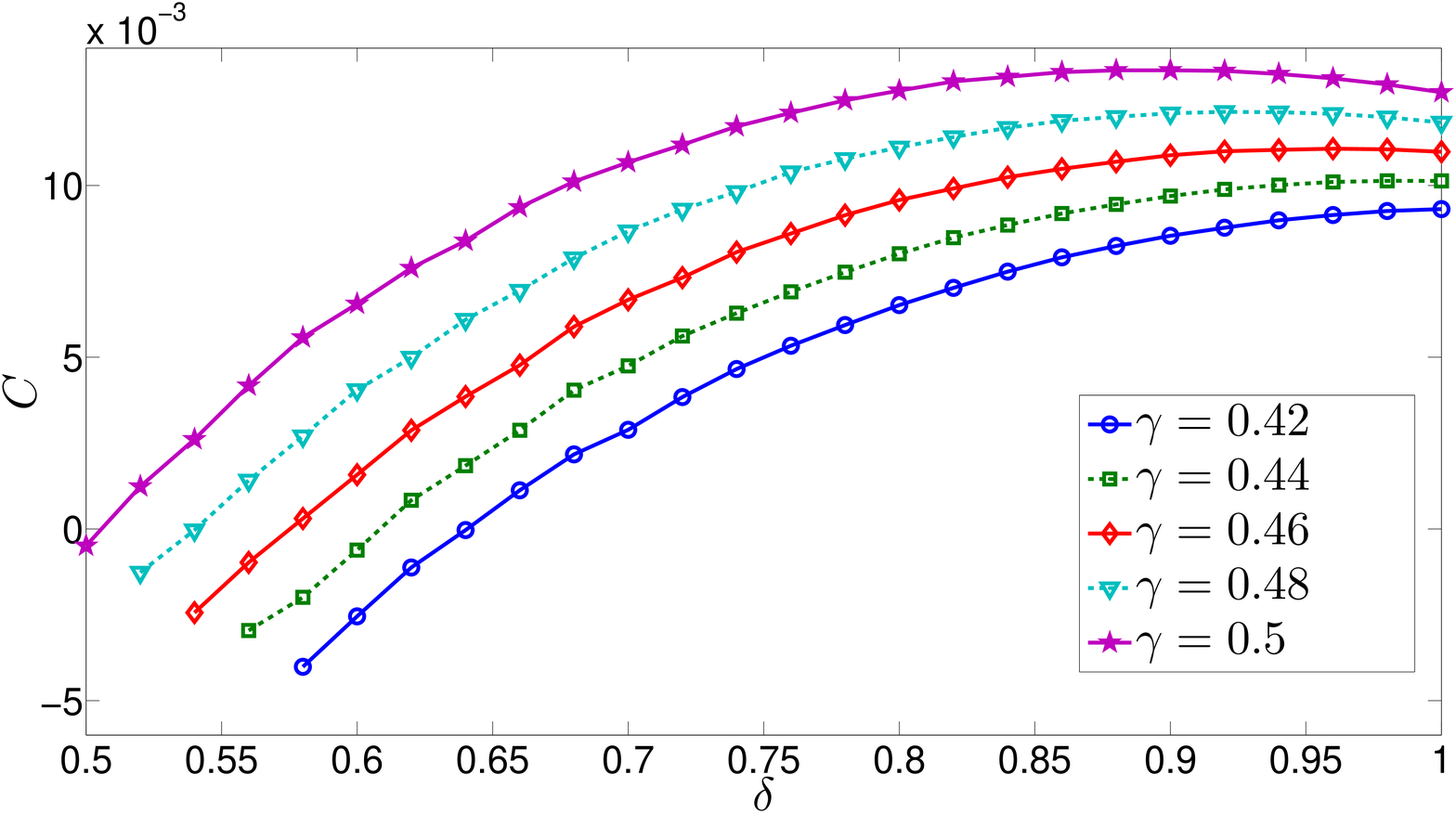}
\includegraphics[width=0.5\textwidth]{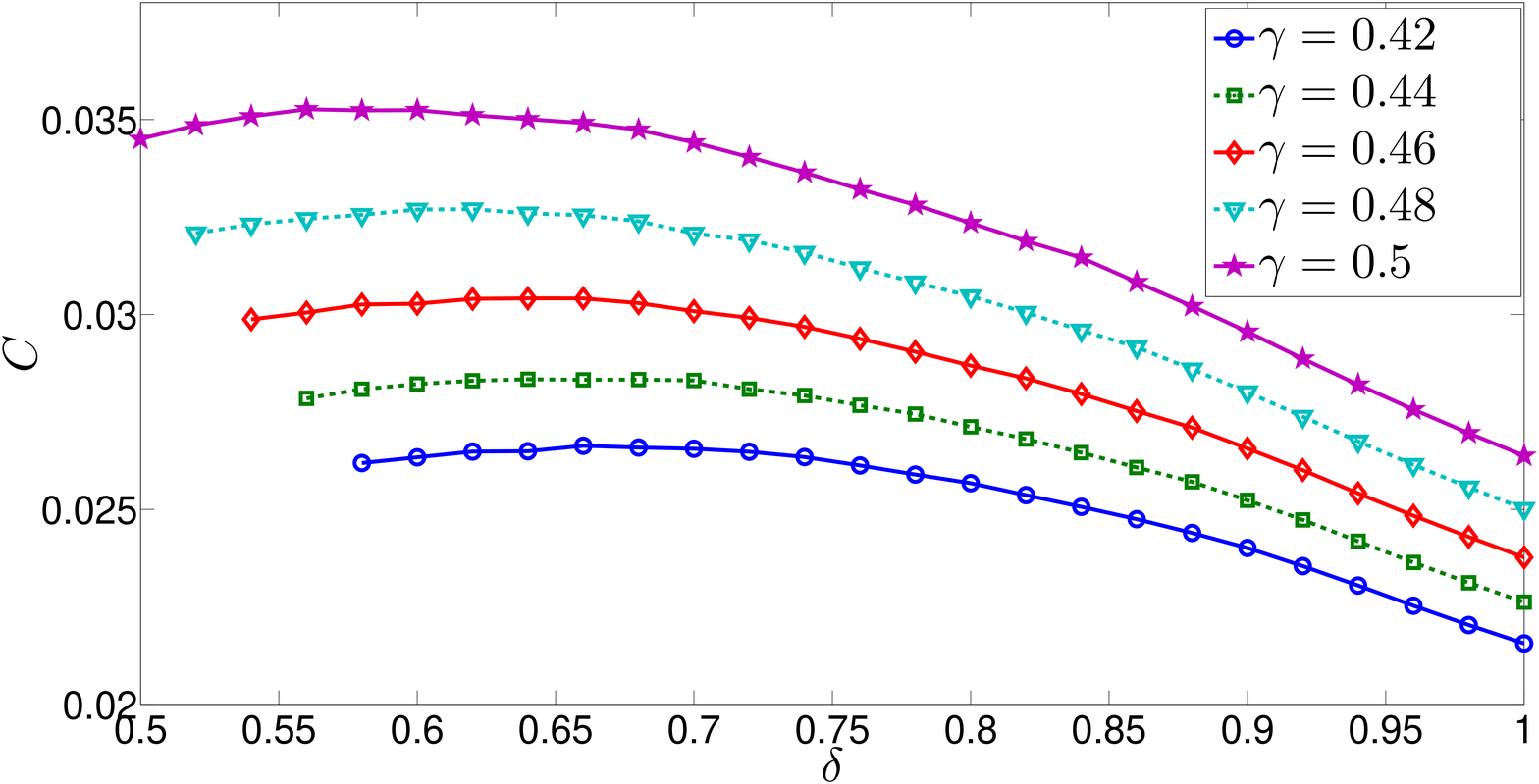}
\end{center}
\vspace*{8pt}
\caption{Cost of the optimal solution given by the SQP algorithm for a buy-program, where $X=0.1$, {\em i.e.} $10\%$ of a market volume and with $N=100$ subintervals, and a concave-convex impact function with $d=0.1$ (top) and $d=1$ (bottom).   We use $1000$ starting points for each optimization and consider only parameters in the no-dynamic-arbitrage region defined by \eqref{eq:2.4}.
         We observe that holding $\gamma$ fixed, expected cost is not a monotonic function of $\delta$.
         For $d=0.1$ negative costs are observed for any value of $\gamma$, while for $d=1$ we observe no negative costs.}  
\label{fig.11} 
\end{figure}

Finally, we extend our investigation to the  whole no-dynamic-arbitrage region, in order to observe how the costs change. Figure \ref{fig.11} shows the expected cost for different values of $\gamma$ as a function of $\delta$, considering separately the case $d=0.1$ and $d=1$. In the former case, we observe a region of parameter space where the cost is negative, while if the convex term is sufficiently large, it is possible to eliminate negative costs in the strong nonlinear region. Further investigation shows that this result does not change as we increase $N$ (data not shown). Finally we note that in the case of a concave-convex impact, as in the case of concave impact, the cost is a non-monotonic function of $\delta$.

In conclusion, our findings emphasize the great importance that the specific shape of the instantaneous market impact function $f$ has for the regularity of the optimal strategy. 

\begin{table}
\small 
\centering
\begin{tabular}{ |l|l|l|l|l|l|}
\hline
 $d$   & $v^{*}$ & $\langle v_{SQP}>0 \rangle$  & $\sigma\left(v_{SQP}>0\right)$ & Cost SQP & Cost VWAP \\
\hline
 $0.1$ & $1.0755$ & $1.1485$ & $0.3193$ & $-0.00245$ & $0.03266$ \\
\hline
 $0.5$ & $0.4256$ & $0.4835$ & $0.0911$ & $0.01674$ & $0.03782$ \\
\hline
 $1$   & $0.2678$ & $0.3229$ & $0.0443$ & $0.02887$ & $0.04428$ \\
\hline
 $2$   & $0.1639$ & $0.2170$ & $0.0292$ & $0.04752$ & $0.05718$ \\
\hline 
\end{tabular}
\caption{In the first column, we report the value of $v^{*}$ for which $f_{G}''\left(v^{*}\right)=0$. The other columns report data regarding the SQP optimization 
         in the case $\gamma=0.45,\,\delta=0.55,\,N=100$. The trading speed decreases when the magnitude of the convex term increases; the cost, instead, increases.  
         The last column report the cost of a VWAP strategy in presence of a concave-convex impact.}  
\label{Table.6}
\end{table}

\section{Conclusions}
\label{Conclusions}

In this paper, we have studied the problem of finding optimal execution strategies in the nonlinear transient impact model of \cite{Gatheral}. We considered several different approaches, discussing their ranges of validity and the characteristics of the solutions they generate, and compared their associated expected costs. We also focused our attention on the presence of different forms of possible price manipulation. 

If the class of admissible strategies is constrained so as to eliminate the possibility of zero trading rates, the cost minimization problem may be cast as an Urysohn integral equation of the first kind.  To solve for the optimal execution strategy with this constraint, we proposed a discretized homotopy analysis technique. The solutions obtained are by construction continuous deformation of initial guesses, such as VWAP. We have shown that the optimal solution found in this way is not time-reversal symmetric, but front-loaded in the case of concave market impact. Our expected cost analysis shows that such solutions substantially outperform conventional execution strategies such as VWAP.  

The integral equation approach can only find local minima of the cost functional. In order to search for a global minimum we therefore perform a direct numerical minimization of the cost function. 
From our analysis using an SQP global optimization technique,  we found that the discretized cost function exhibits a rugged landscape, with many local minima separated by peaks. The global minimum (as well as other many other suboptimal minima) corresponds to a strategy that alternate buys and sells, and thus allows for transaction-triggered price manipulation. In particular, for a buy program the optimal solution consists of short bursts of trading at high rates separated by long periods of selling at low trading rates. Even worse, when the nonlinearity is strong and when the discretization is sufficiently fine, we find  negative execution costs. Coupled with the transient property of market impact in our model, this implies that the nonlinear transient impact model with purely power-law concave impact admits price manipulation, {\em i.e.} round trips with positive expected revenues, eliminating such a model from consideration for practical use.  In particular, this shows that the conditions \eqref{eq:2.4} of \cite{Gatheral} for no price manipulation are necessary but not sufficient.  

We also studied optimal monotonic strategies where selling(buying) is disallowed in a buy(sell) program. This optimization is performed by a derivative-free direct search method, specifically the generating set search algorithm.  In this case, we eliminate transaction-triggered price manipulation, we no longer see negative expected costs, and the corresponding optimal strategy is sparse. In other words, it is optimal to trade in a few intense bursts separated by long periods of no trading.

Finally, we propose two ways to regularize the transient impact model, both of them reflecting important features of the market that are neglected in the simple version of the model.  In the first approach, we add a spread cost, effectively performing a $L_1$ or LASSO regularization, which discourages negative trades for buy programs. In the presence of sufficiently strong transaction costs, the optimal execution strategy no longer exhibits a negative expected cost, and the optimal buy strategy becomes sparse and bursty, with a few moments of strong positive trading separated by periods of zero or slightly negative trading. In the second approach, we add to the market impact function a convex component which makes it very costly to trade at high trading rates. We observe that if the convex component is sufficiently large, negative expected execution costs are again eliminated. 

Our results indicate that a fully satisfactory model of market impact, which is free of price manipulation and consistent with empirical observation is still lacking, leaving a fruitful field for future research \cite{Donier}. 

\section*{Acknowledgments}
We thank Dario Trevisan for useful discussions regarding calculus of variations. This research was partially supported by HSBC Holdings plc.

\appendix

\section{Appendices}
\label{Appendices}

\subsection{Dang's fixed point algorithm}\label{sec:DangFixedPoint}
\label{Fixed point method}
In this section we analyze an iteration scheme proposed by Dang \cite{Dang} to find a numerical solution of equation \eqref{eq:3.24} by quadrature methods. Dang considers a power law impact function
$f\left(v\right)=sign\left(v\right) |v|^{\delta}$ where $\delta >1$. If we want to study the case $\delta <1$ we should find a way to handle the problem 
of an infinite first derivative of $f'\left(v\right)$ for $v=0$. We define a perturbative version of the market impact function $\hat{f}\left(v\right)$
\begin{eqnarray}
 \label{eq:A.2.1}
\ \hat{f}\left(v\right)&=& sign \left(v\right) \left(\epsilon+|v|\right)^{\delta} \nonumber\\
\ \hat{f}'\left(v\right)&=& \delta \left(\epsilon+|v|\right)^{\delta-1}
\end{eqnarray}
where the perturbation is given by the parameter $0<\epsilon\ll 1$. In this way the trading rate can assume any value on $\mathbb{R}$ 
We search a solution $v_{i}\left(\epsilon\right)$ of the equation \eqref{eq:3.24} with a perturbation near to zero, {\em i.e.} $\epsilon \approx 0$.

Dang defines a nonlinear map from equation \eqref{eq:3.24} and searches a fixed point for the map\footnote[7]{We have used different discretization methods to approximate the integral equation, including the product Nystrom method \cite{Press} used by Dang. We find that our conclusions do not depend on the kind of discretization used.} starting from the initial guess $v_{i}^{0}$. This means that we have 
a sequence of approximations for the solution $v_{i}^{0},v_{i}^{1},v_{i}^{2},\cdots$, the convergence to a fixed point of the map is a necessary condition in order
to have a convergent iteration scheme \cite{Berger,Browder}. 
The iterative scheme is defined by a Taylor expansion 
\begin{equation}
 \label{eq:A.2.2}
F_{ij}\left(v^{m}\right)\approx F_{ij}\left(v^{m-1}\right)+
\left(v_{j}^{m}-v_{j}^{m-1}\right)F'_{ij}\left(v^{m-1}\right)\,,
\end{equation}
where
\begin{equation}
 \label{eq:A.2.3}
F'_{ij}\left(v\right)=\begin{cases} \hat{f}'\left(v_{j}\right), & \mbox{if }j\leq i \\ \hat{f}'\left(v_{i}\right), & \mbox{if }j>i
\end{cases}
\end{equation}
is the derivative\footnote[8]{Here we can observe the fundamental difference between our DHAM approach and Dang's fixed point approach. The DHAM approach is based on the observation that this dynamical system 
has to be regarded as a bi-dimensional system; in contrast, Dang develops a method that lacks interaction between past and present values of the trading rate $v$.} of $F_{ij}(v)$ in equation \eqref{eq:3.25} with respect to $v_{j}$. Following Dang, we solve for each $m$-th iteration the following linear system:
\begin{equation}
 \label{eq:A.2.4}
 \boldsymbol{K} v^{m}= \tilde{c}.
\end{equation}
This defines a nonlinear map $M$
\begin{equation}
 \label{eq:A.2.5}
v^{m}=\boldsymbol{K}^{-1}\left(v^{m-1}\right)\tilde{c}\left(v^{m-1}\right)=M\left(v^{m-1}\right)\,,
\end{equation}
where
\begin{eqnarray}
 \label{eq:A.2.6}
\ K_{ij}&=&G_{ij}F'_{ij}\left(v^{m-1}\right)\,, \nonumber\\
\ \tilde{c}_{i}&=&\lambda-\sum_{j=1}^{i}G_{ij}\,\left(F_{ij}\left(v^{m-1}\right)-v_{j}^{m-1}F'_{ij}\left(v^{m-1}\right)\right).
\end{eqnarray}
We search a solution that is a fixed point of the map $M$, {\em i.e.} $v^{*}=M\left(v^{*}\right)$, as in Dang \cite{Dang}. Starting from the initial guess $v^{0}$ we look for a value $\bar{m}$
for which the $N$-dimensional vector $v^{m}$ is a constant vector for $m> \bar{m}$. A simple scalar quantity controlling if the fixed point has been reached  is the mean field, defined as
$\bar{v}^{m}=1/N\,\sum_{i=1}^{N}\,v_{i}^{m}$. The mean field $\bar{v}$ is also useful in order to implement a procedure that leads to determine the value of $\lambda$ that satisfies the constraint \eqref{eq:3.29} on the total quantity traded. 
It is clear that for a general dynamic system the condition of a constant mean field $\bar{v}$ does not guarantee the existence of a fixed point since there can be also more complex attractors, {\em e.g.}
a chaotic attractor, where the mean field is constant. Here we are interested in studying only a necessary condition in order to find a fixed point of the nonlinear map $M$. Moreover, it is important to notice that the evolution of the map $M$ does not satisfy the constraint of equation \eqref{eq:3.29}, {\em i.e.}
if the initial point $v^{0}$ is on the hyper-plane, it could go outside of it when we iterate the map $M$. This is the reason for which we should tune the constant $\lambda$ in order to satisfy
our constraint within some precision.

Dang's proposal for the initial guess $v_{i}^{0}=v$, $\forall i$, is a constant vector whose components are given by  
\begin{equation}
 \label{eq:A.2.7}
vf'\left(v\right)\sum_{j=1}^{N}\,G_{1j}=\lambda.
\end{equation}
In principle the fixed point could depend on the starting point $v^{0}$ that we choose as initial guess for the solution.
We add to this initial guess a set of initial guesses uniformly distributed on the simplex to investigate the presence of different attractors. The convergence of this method is determined by two parameters: $N$ and $\delta$.

We performed an extensive analysis of the convergence of the Dang's fixed point method. The convergence criterion is that the relative standard deviation of the mean field is below some threshold, typically $10^{-9}$. The result of this analysis for $\gamma=0.5$ and $\epsilon=0$ is shown in Figure \ref{fig:Conv_Region&Squared_Error}, but similar results are obtained for small values of $\epsilon$. We observe that when
$N$ increases, the set of values of $\delta$ for which we have the convergence decreases, {\em i.e.} we have a $\delta_{min}$ under which the procedure does not converge, {\em i.e.} we do not find a fixed point for the map $M$, see Figure \ref{fig:Conv_Region&Squared_Error}. When the procedure converges, the solution does not depend from the initial guess, i.e the map $M$ has only one attractive basin. Moreover, the convergence is
extremely fast, reaching the fixed point in very few iterations of the map. As $N$ increases however, Dang's algorithm converges only in the weakly nonlinear case. 

\begin{figure}[t]
\begin{center}
\includegraphics[width=0.9\textwidth]{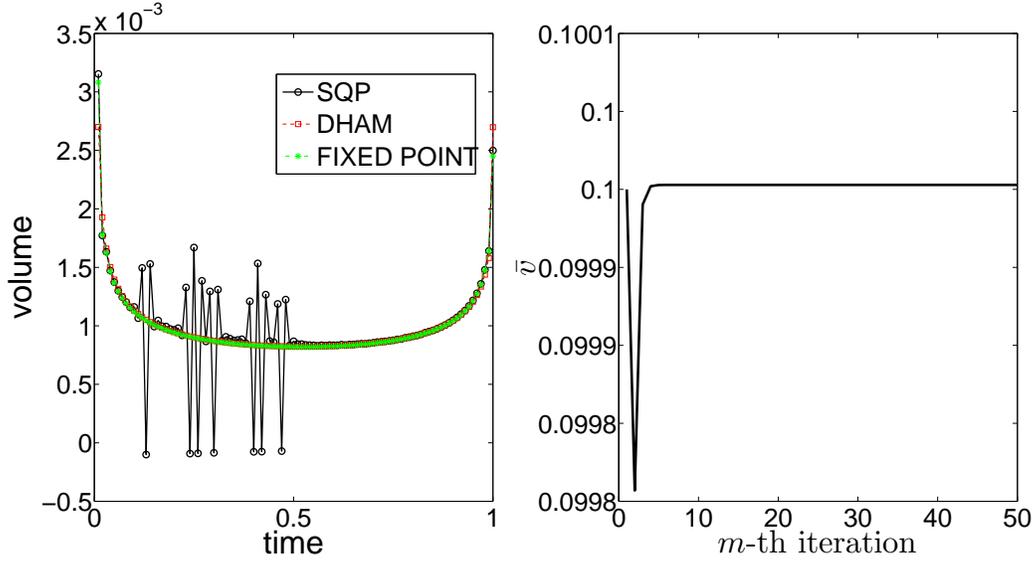}
\end{center}
\vspace*{8pt}
\caption{Left panel. Optimal solutions using the Dang fixed point method, the DHAM and the SQP method in the weakly nonlinear regime: $\gamma=0.5,\delta=0.95,N=100,T=1,X=0.1$. The right panel reports the fast convergence of the mean field $\bar{v}^{m}$ of the fixed point method with a VWAP initial guess and $\lambda=2.87*10^{-3}$.
         Notice that the initial guess goes outside the constraint at the beginning of the iteration procedure. }
\label{fig.A.2}
\end{figure}

Confining our attention to a regime where Dang's algorithm does converge, how does the solution achieved with Dang's algorithm compare with the solution achieved using the  DHAM and SQP methods we propose?  In Figure \ref{fig.A.2} we report results for the very weakly nonlinear case with $N=100,T=1,X=0.1,\gamma=0.5,\delta=0.95,\epsilon=10^{-6}$.  We find that in this case the Dang method converges very quickly (right panel) and the solution is essentially coincident with the one obtained with the DHAM method (left panel). The solution of the numerical optimization using the SQP algorithm is  oscillating, alternating between positive and negative trading rates. 

In conclusion, Dang's algorithm converges only under weak nonlinearity and/or when discretization of the time interval is very coarse. When the Dang method does converge, the solution is very close to the one obtained with the DHAM method proposed in this paper.  However the DHAM approach is applicable not only in the weakly nonlinear regime where Dang's algorithm converges, but also in the strongly nonlinear regime.

\subsection{Homotopy derivatives}\label{appHomo}

As explained in Section \ref{sec:Homotopy}, the so-called homotopy derivative is used to deduce deformation equations of equation \eqref{eq:3.8} for any order greater than one, {\em i.e.} to compute $v^{m}$ for $m>1$. 
Such computations are difficult because they depend on the nonlinear operator $\mathcal{N}$. 
For our problem, we need to compute the homotopy derivative of a power law function. 

A nonlinear power law function with a integer exponent, {\em i.e.} $f\left(\phi\right)=\phi^{k},\,\,k\in \mathbb{N}$, was studied by Molabahrami and Khani \cite{Molabahrami}
to find an approximate solution of the Burgers-Huxley equation. The case of a real power-law index was studied by Wang et al. \cite{Wang} analyzing the flow of a power-law fluid film on an unsteady stretching surface.
Our case \eqref{eq:2.9} is more complicated than the case of a simple power-law function. Recent results given by Turkyilmazoglu \cite{Turkyilmazoglu} and Liao \cite{Liao} show how to compute
the homotopy-derivative of any smooth function. The homotopy-derivative $\mathcal{D}^{m}\left[f\left(\phi\right)\right]=\left(1/m!\right)\partial^{m}f\left(\phi\right)/\partial p^{m}$ is given by the recursive relation
\begin{equation}
 \label{eq:3.12}
\mathcal{D}^{m}\left[f\left(\phi\right)\right]=\sum_{k=0}^{m-1}\left(1-\frac{k}{m}\right)\mathcal{D}^{m-k}\left[\phi\right]\frac{\partial}{\partial \phi}\left\lbrace \mathcal{D}^{k}\left[f\left(\phi\right)\right] \right\rbrace,
\end{equation}
evaluated at $p=0$.  
The sum consists of  two homotopy-derivative terms. The first gives $v^{m-k}$, the second term gives polynomial terms of $v^{i}$ that multiply $f^{i}\left(v^{0}\right),\,i=1,\cdots,m$, derivatives of $f$ evaluated at the initial guess.  If the market impact function $f$ is of the form $f(v)\propto v^\delta$ with $\delta <1$, all such derivatives diverge at $v=0$. As mentioned in Section \ref{sec:Homotopy}, we avoid this problem by choosing the initial guess to be a strictly positive (or negative) function of time.  
Using the $m-1$-th order of equation \eqref{eq:3.12} in equation \eqref{eq:3.9} we can express the $m$-th homotopy derivative $v^{m}$ as a complicated function of the previous $m-1$ derivatives.

To handle the nonlinearity of equation \eqref{eq:2.9} in the HAM framework we need a further step. The coupling between past and future values of trading rates implies that we have to consider our problem as a two-dimensional system. This means that we
use the homotopy-derivative for systems described by two variables \cite{Liao}, for example $u$ and $w$, in which we have a nonlinear coupling between them given by $f\left(u,w\right)$
\begin{equation}
 \label{eq:3.13}
\mathcal{D}^{m}\left[f\left(\phi,\psi\right)\right]=\sum_{k=0}^{m-1} \left(1-\frac{k}{m}\right)u^{m-k}\mathcal{D}^{k}\left[\frac{\partial f\left(\phi,\psi\right)}{\partial \phi}\right]+
                                                    \left(1-\frac{k}{m}\right)w^{m-k}\mathcal{D}^{k}\left[\frac{\partial f\left(\phi,\psi\right)}{\partial \psi}\right],
\end{equation}
where $\phi\,,\psi$ are the Maclaurin series of $u\,,w$ respectively. Thus, given the expression of equation \eqref{eq:2.9} as $F\left(v\left(s\right),v\left(t\right)\right)$, we are able to apply the HAM to our optimal
execution problem using the equation \eqref{eq:3.13} and choosing a function with a given sign as initial guess.


\medskip

\begin{thebibliography}{00}


\bibitem{Abbasbandy2} Abbasbandy, S., Shivanian, E., Vajravelu, K., Mathematical properties of $\hslash$-curve in the framework of the homotopy analysis method, {\it Commun. Nonlinear Sci. Numer. Simulat.} {\bf16} (2011) 4268--4275.
\bibitem{Abbasbandy5} Abbasbandy, S., Homotopy analysis method for the Kawahara equation, {\it Nonlinear Analysis: Real World Applications} {\bf11} (2010) 307--312.  
\bibitem{Abergel} Abergel F., Bouchaud J.-P., Foucault T., Lehalle C.-A., Rosenbaum M., \textit{Market microstructure: confronting many viewpoints} (The Wiley finance series, Padstow, Cornwall, UK, 2012).
\bibitem{Akyildiz} Akyildiz, F. T., Vajravelu, K., Magnetohydrodynamic flow of a viscoelastic fluid, {\it Physics Letters A} {\bf372} (2008) 3380--3384.
\bibitem{Alfonsi2} Alfonsi, A., and Schied, A., Capacitary measures for completely monotone kernels via singular control, {\it SIAM J. CONTROL OPTIM.} {\bf51} (2013) 1758--1780.
\bibitem{Alfonsi3} Alfonsi, A., Schied, A., and Slynko, A., Order book resilience, price manipulation, and the positive portfolio problem, Preprint available at SSRN (2009).
\bibitem{Almgren1} Almgren, R., and Chriss, N., Value under liquidation, {\it Risk} {\bf12} (1999) 61--63.
\bibitem{Almgren2} Almgren, R., and Chriss, N., Optimal Execution of Portfolio Transactions, {\it J. Risk} {\bf3} (2000) 5--39.
\bibitem{Almgren3} Almgren, R., Optimal Execution with Nonlinear Impact Functions and Trading-Enhanced Risk, {\it Appl. Math. Finance} {\bf10} (2003) 1--18. 
\bibitem{Bacry} Bacry, E., Iuga, A., Lasnier, M., and Lehalle, C.-A., Market impacts and the life cycle of investors orders, \url{http://arxiv.org/abs/1412.0217} (2014).



\bibitem{Berger} Berger, M. S., Nonlinearity and functional analysis, Lectures on Nonlinear Problems in Mathematical Analysis. Academic Press, Inc.(London) Ltd(1977).
\bibitem{Bertsimas} Bertsimas, D., and Lo, A., Optimal Control of Execution Costs, {\it Journal of Financial Markets}  {\bf1} (1998) 1--50.
\bibitem{Bouchaud2} Bouchaud, J.-P., Farmer, J.D. and Lillo, F., How markets slowly digest changes in supply and demand. In {\it Handbook of Financial Markets: Dynamics and Evolution}, edited by T.Hens and K. Schenk-Hoppe, (2008), Academic Press: New York.
\bibitem{Bouchaud3} Bouchaud, J.-P., Kockelkoren, J., Potters, M., Random walks, liquidity molasses and critical response in financial markets, {\it Quantitative Finance} {\bf6} (2006) 115--123.
\bibitem{Bouchaud} Bouchaud, J.-P., Gefen, Y., Potters, M. and Wyart, M., Fluctuations and response in financial markets: the subtle nature of 'random' price changes, {\it Quantitative Finance} {\bf4} (2004) 176--190.
\bibitem{Brodie} Brodie, J., Daubechies, I., De Mol, C., Giannone, D., and Loris, I., Sparse and stable Markowitz portfolios, {\it PNAS} {\bf106} (2009) 12267--12272.
\bibitem{Browder} Browder, F.E., Construction of Fixed Points of Nonlinear Mappings in Hilbert Space, {\it Journal of Mathematical Analysis and Applications } {\bf20} (1967) 197--228.
\bibitem{Brown} Brown, K. S. and Sethna, J. P., Statistical mechanical approaches to models with many poorly known parameters, {\it Physical Review E} {\bf68} (2003) 021904-1--021904-9. 
\bibitem{Busseti} Busseti, E., and Lillo, F., Calibration of optimal execution of financial transactions in the presence of transient market impact, {\it J. Stat. Mech.} {\bf} (2012) P09010.
\bibitem{Chiang} Chiang, A. C.. Fundamental methods of mathematical economics. McGraw-Hill, Inc, (1984).
\bibitem{Dang} Dang, N. M., Optimal execution with transient impact, {\it } {\bf} (2014) SSRN-id2183685.
\bibitem{Donier} Donier, J., Bonart, J., Mastromatteo, I., and Bouchaud, J.-P., A fully consistent, minimal model for non-linear market impact, \url{http://arxiv.org/abs/1412.0141v2} (2014). 
\bibitem{Gatheral2} Gatheral, J., Schied, A. and Slynko, A., Transient linear price impact and Fredholm integral equations, {\it Mathematical Finance} {\bf22} (2012) 445--474.
\bibitem{GatheralSchiedSlynkoEcon} Gatheral, J., Schied, A. and Slynko, A., Exponential resilience and decay of market impact, {\em Econophysics of Order-driven Markets}, Springer, Berlin, 225--236 (2011).
\bibitem{GatheralSchied} Gatheral, J., Schied, A., Dynamical models of market impact and algorithms for order execution, in {\it Handbook of Systemic Risk} (J.-P. Fouque and J. A. Langsam editors), Cambridge University Press 2013.
\bibitem{Gatheral} Gatheral, J., No-dynamic-arbitrage and market impact, {\it Quantitative Finance} {\bf10} (2010) 749--759.
\bibitem{Hetmaniok} Hetmaniok, E., S\l{}ota, D., Trawi\'{n}ski, T., Witu\l{}a, R., Usage of the homotopy analysis method for solving the nonlinear and linear integral equations of the second kind, {\it Numer. Algor.} {\bf67} (2014) 163--185.
\bibitem{Kolda2} Kolda, T. G., Lewis, R. M. and Torczon, V., Stationarity results for generating set search for linearly constrained optimization, {\it SIAM Journal on Optimization} {\bf17} (2006) 943--968. 
\bibitem{Kolda} Kolda, T. G., Lewis, R. M. and Torczon, V., Optimization by Direct Search: New Perspectives on Some Classical and Modern Methods, {\it SIAM Review} {\bf45} (2003) 385--482.
\bibitem{Liao2} Liao, S., On the homotopy analysis method for nonlinear problems, {\it Applied Mathematics and Computation} {\bf147} (2004) 499--513.
\bibitem{Liao3} Liao, S., Notes on the homotopy analysis method: Some definitions and theorems, {\it Commun. Nonlinear Sci Numer. Simulat.} {\bf14} (2009) 983--997.
\bibitem{Liao4} Liao, S., Beyond Perturbation: Introduction to the Homotopy Analysis Method. Boca Raton: Chapman Hall CRC/Press; (2003). 
\bibitem{Liao} Liao, S., Homotopy Analysis Method in Nonlinear Differential Equations. Springer (2012).
\bibitem{Lillo} Lillo, F., Farmer, J.D. and Mantegna, R. N., Master curve for price-impact function, {\it Nature} {\bf421} (2003) 129--130.  
\bibitem{Liu} Liu, C.-S., The essence of the generalized Taylor theorem as the foundation of the homotopy analysis method, {\it Commun. Nonlinear Sci. Numer. Simulat.} {\bf16} (2011) 1254--1262.  
\bibitem{Magnus} Magnus, J. R., Neudecker, H.. Matrix Differential Calculus with Applications in Statistics and Econometrics, 3rd Edition. John Wiley and \& Sons Ltd, (2007). 
\bibitem{Matlab1} Optimization Toolbox User's Guide, 1990-2012 by The MathWorks, Inc.
\bibitem{Matlab2} Global Optimization Toolbox User's Guide, 2004-2012 by The MathWorks, Inc.    
\bibitem{Matlab3} Parallel Computing Toolbox User's Guide, 2004-2012 by The MathWorks, Inc.
\bibitem{Molabahrami} Molabahrami, A., Khani, F., The homotopy analysis method to solve the Burgers-Huxley equation, {\it Nonlinear Analysis: Real World Applications} {\bf10} (2009) 589--600.
\bibitem{Obizhaeva} Obizhaeva, A. A., Wang, J., Optimal trading strategy and supply/demand dynamics, {\it Journal of Financial Markets} {\bf16} (2013) 1--32.
\bibitem{Polyanin} Polyanin, A. D., Manzhirov, A., Handbook of integral equations, Second edition. Chapman \& Hall/CRC(2008). 
\bibitem{Predoiu} Predoiu, S., Shaikhet, G., Shreve, S, Optimal execution in a general one-sided limit-order book, {\it SIAM Journal on Finance Mathematics} {\bf2} (2011) 183--212. 
\bibitem{Press} Press, W.H., Teukolsky, S.A., Vetterling, W.T., Flannery, B.P., Numerical Recipes 3rd Edition: The Art of Scientific Computing, Cambridge University Press, 3rd Edition (2007).
\bibitem{Sanchez2} S\'{a}nchez, D. A., Calculus of variations for integrals depending on a convolution product, {\it Annali della Scuola Normale Superiore di Pisa, Classe di Scienze } (3) {\bf18} (1964) 233--254.
\bibitem{Sanchez3} S\'{a}nchez, D. A., Some existence theorems in the calculus of variations, {\it Pacific Journal of Mathematics} {\bf19} (1966) 357--363.
\bibitem{Sanchez} S\'{a}nchez, D. A., On Extremals for Integrals Depending on a Convolution Product, {\it Journal of Mathematical Analysis and Applications} {\bf11} (1965) 213--216.
\bibitem{Turkyilmazoglu} Turkyilmazoglu, M., A note on the homotopy analysis method, {\it Applied Mathematics Letters} {\bf23} (2010) 1226--1230.
\bibitem{Verma} Verma, A., Schug, A., Lee, K. H., Wenzel, W., Basin hopping simulations for all-atom protein folding, {\it The Journal of Chemical Physics} {\bf124} (2006) 044515.
\bibitem{Wang} Wang, C., Pop, I., Analysis of the flow of a power law fluid film on an unsteady stretching surface by means of the homotopy analysis method, {\it J. Non-Newtonian Fluid Mech.} {\bf138} (2006) 161--172.
\bibitem{Waterfall} Waterfall, J. J., Casey, F. P., Gutenkunst, R. N., Brown, K. S., Myers, C. R., Brouwer, P. W., Elser, V., and Sethna, J. P., Sloppy-Model Universality Class and the Vandermonde Matrix, {\it Physical Review Letters} {\bf97} (2006) 150601-1--150601-4.
\bibitem{Wazwaz} Wazwaz A.-M., Linear and Nonlinear Integral Equations, methods and applications. Springer-Verlag (2011).
\bibitem{Weinberger2} Weinberger, E., Correlated and Uncorrelated Fitness Landscapes and How to Tell the Difference, {\it Biol. Cybern.} {\bf63} (1990) 325-336.
\bibitem{Weinberger} Weinberger, E. D., Local properties of Kauffman's $N$-$k$ model: A tunably rugged energy landscape, {\it Physical Review A} {\bf44} (1991) 6399-6413.
\bibitem{kyle} Kyle, A. S., Continuous auctions and insider trading, {\it Econometrica } (1985) 1315--1335.
\bibitem{Zarinelli} Zarinelli, E., Treccani, M., Farmer, J. D., and Lillo, F., Beyond the square root: Evidence for logarithmic dependence of market impact on size and participation rate, \url{http://arxiv.org/abs/1412.2152v1} (2014). 

\end{thebibliography}
  \end{document}